\documentclass[aps,prd,preprintnumbers,superscriptaddress,nofootinbib,amsmath,amssymb,showpacs,preprint]{revtex4}
\pdfoutput=1
\usepackage{graphicx}
\usepackage{dcolumn}
\usepackage{bm}
\usepackage{color}

\usepackage{slashed}

\input{colordvi.tex}

\newcommand{\beq}{\begin{equation}}
\newcommand{\eeq}{\end{equation}}
\newcommand{\beqa}{\begin{eqnarray}}
\newcommand{\eeqa}{\end{eqnarray}}

\begin{document}

\begin{flushright}
CTPU-PTC-19-14 \\
RESCEU-3/19 
\end{flushright}

\title{Magnetogenesis from a rotating scalar: \\  \`a la scalar chiral magnetic effect}

\author{Kohei Kamada}
\email[Email: ]{kohei.kamada"at"resceu.s.u-tokyo.ac.jp}
\affiliation{Research Center for the Early Universe (RESCEU),
Graduate School of Science, The University of Tokyo, Tokyo 113-0033, Japan}
\affiliation{Center for Theoretical Physics of the Universe, Institute for Basic Science (IBS), Daejeon 34126, Korea}

\author{Chang Sub Shin}
\email[Email: ]{csshin"at"ibs.re.kr}
\affiliation{Center for Theoretical Physics of the Universe, Institute for Basic Science (IBS), Daejeon 34126, Korea}


\begin{abstract}
The chiral magnetic effect (CME) is a phenomenon in which an electric current is induced parallel to an external magnetic field in the presence of chiral asymmetry in a fermionic system. In this paper, we show that the electric current induced by the dynamics of a pseudo-scalar field which anomalously couples to electromagnetic fields can be interpreted as closely analogous to the CME. In particular, the velocity of the pseudo-scalar field, which is the phase of a complex scalar, indicates that the system carries a global U(1) number asymmetry as the source of the induced current. We demonstrate that 
an initial kick to the phase-field velocity and an anomalous coupling between the phase-field and gauge fields are naturally provided, in a set-up such as the Affleck-Dine mechanism. The resulting asymmetry carried by the Affleck-Dine field can give rise to instability in the (electro)magnetic field. Cosmological consequences of this mechanism are also investigated. \end{abstract}
\maketitle

\section{introduction} 

In the chiral magnetic effect (CME)~\cite{Vilenkin:1980fu,Fukushima:2008xe}, electric currents are induced by magnetic fields in the presence of chiral asymmetry. 
Since the CME originates from quantum anomalies~\cite{Adler:1969gk,Bell:1969ts}, which are ubiquitous in quantum systems regardless of their energy scales, 
it can play an important role in a variety of settings: relativistic heavy-ion collisions~\cite{Fukushima:2008xe,Kharzeev:2010gd,Burnier:2011bf,Hongo:2013cqa,Yee:2013cya,Hirono:2014oda,Huang:2015oca,Kharzeev:2015znc,Shi:2017cpu}, Weyl semimetals~\cite{Zyuzin:2012tv,Goswami:2012db,Chen:2013mea,Basar:2013iaa,Hosur:2013kxa,Landsteiner:2013sja,Chernodub:2013kya}, 
astrophysical objects such as neutron stars~\cite{Charbonneau:2009ax,Ohnishi:2014uea,Grabowska:2014efa,Kaminski:2014jda,Sigl:2015xva}, supernovae~\cite{Yamamoto:2015gzz,Masada:2018swb}, etc. 
Moreover, it has been argued that in the early Universe, when the chiral asymmetry was well preserved~\cite{Campbell:1992jd},  the CME can cause a tachyonic instability in the (hyper)magnetic fields~\cite{Joyce:1997uy,Tashiro:2012mf}. 
This ``chiral plasma instability" has recently been studied with full magnetohydrodynamic simulations~\cite{Rogachevskii:2017uyc,Brandenburg:2017rcb,Schober:2017cdw,Schober:2018ojn},\footnote{See also Ref.~\cite{Figueroa:2019jsi} for the lattice study on the chiral plasma instability.} which showed that a maximal transfer of chiral asymmetry to magnetic helicity is likely to occur. 
One implication is that these maximally helical (hyper)magnetic fields 
may be the source of the baryon asymmetry of the Universe~\cite{Giovannini:1997gp,Giovannini:1997eg,Bamba:2006km,Fujita:2016igl,Kamada:2016eeb,Kamada:2016cnb,Kamada:2018tcs}.\footnote{This mechanism can be used to explain the proposed intergalactic magnetic fields from blazar observations~\cite{Neronov:1900zz,Tavecchio:2010mk,Dolag:2010ni,Essey:2010nd,Taylor:2011bn,Takahashi:2013lba,Finke:2015ona,Biteau:2018tmv}. However, since baryons may be overproduced in this case~\cite{Kamada:2016cnb,Kamada:2018tcs}, more careful analysis is required around the period of electroweak symmetry breaking.}

In this scenario the chiral asymmetry, the origin of the CME, is usually carried by light fermions.  On the other hand, even if there are no light fermions, 
such an effect can also be observed in the low-energy effective theory of axions~\cite{Peccei:1977hh,Peccei:1977ur,Weinberg:1977ma,Wilczek:1977pj}, where it takes the form of an anomalous coupling between the axion and gauge bosons~\cite{Srednicki:1985xd}.\footnote{See also the recent discussions in Ref.~\cite{Quevillon:2019zrd}.}  In that case the background dynamics of the axion field also induces an electric current, similar to the CME.  
Accordingly, it has been argued that the chiral asymmetry can be interpreted  as an axion-like scalar degree of freedom~\cite{Frohlich:2000en,Frohlich:2002fg,Boyarsky:2015faa}. 
Indeed, the cosmological coherent dynamics of axion-like fields  
has been utilized to show how cosmological magnetic fields 
could be generated during inflation~\cite{Turner:1987bw,Garretson:1992vt,Anber:2006xt,Domcke:2018eki} and after inflation~\cite{Adshead:2015pva,Fujita:2015iga,Adshead:2016iae,Cuissa:2018oiw}, 
or later times~\cite{Choi:2018dqr}.  

In this paper, we more directly investigate the analogy between the magnetic field amplification from the axion-like field and that from the CME. 
In particular, we find that the dynamics of the axion-like field gives rise to a non-vanishing chemical potential for the global U(1)$_\mathrm{PQ}$ symmetry,\footnote{
We use the notation U(1)$_\mathrm{PQ}$ to denote a global U(1) symmetry 
which is broken anomalously by gauge interactions, as first proposed by Peccei and Quinn.} similar to the chiral chemical potential. 
The difference between these two scenarios lies in the conservation of the chirality/global U(1)$_\mathrm{PQ}$ charge. 
In the CME case, after chiral asymmetry is generated, chiral symmetry is approximately restored so that a maximal transfer of chiral asymmetry to magnetic helicity is possible.  In contrast, in the axion case, the full dynamics is driven by a scalar potential which explicitly breaks U(1)$_\mathrm{PQ}$ symmetry. Hence, the U(1)$_\mathrm{PQ}$ charge is not conserved when the magnetic fields are amplified. From this perspective, the magnetic field amplification from axion-like fields reported in the literatures is not maximally efficient. 

Notably, the anomalous coupling of  axion-like fields is not limited to QCD or string-theoretical axions but is common to pseudo-scalar fields in general. 
One example in cosmology that takes advantage of the pseudo-scalar dynamics is the Affleck-Dine (AD) mechanism for baryogenesis~\cite{Affleck:1984fy,Dine:1995kz}.
In this mechanism, a complex scalar field with baryonic (leptonic) charge acquires a large expectation value in the early Universe, and a non-vanishing velocity in the phase direction (i.e. the Nambu-Goldstone mode) 
is provided by the explicit baryon (lepton) number-violating interactions. Baryon asymmetry is also generated at the same time. 
In implementations of the AD mechanism, the baryon (lepton) number-violating interactions quickly become ineffective, such that a baryon (lepton) number is conserved after its generation. The interesting consequence is that the complex scalar exhibits a coherent rotation in the field space with a constant angular velocity. 

In this paper, we point out that if a complex scalar field of the AD mechanism is charged under the U(1)$_\mathrm{PQ}$,  
maximally helical magnetic fields can be easily obtained. 
Thanks to the anomalous coupling of the phase-field to the U(1) gauge fields 
and a conserved current  induced by a  {\it constant} U(1)$_\mathrm{PQ}$ asymmetry, only one helicity mode of magnetic fields is amplified. 
 Thus, our mechanism is more efficient than previous realizations of magnetogenesis through the axion-like field dynamics
and is similar to the chiral plasma instability. 

This idea has several important implications for model building in early Universe cosmology. 
First of all, if axion-like fields such as Peccei-Quinn-Weinberg-Wilczek (PQWW)  axions~\cite{Peccei:1977hh,Peccei:1977ur,Weinberg:1977ma,Wilczek:1977pj} experience cosmological evolution like the AD mechanism, it leads to a new mechanism of magnetogenesis. We also note that such magnetic field amplification can even occur in the usual AD mechanism, i.e., even in the minimal supersymmetric standard model (MSSM) and other supersymmetric extensions of the Standard Model of particle physics (SM).  Indeed, we shall show that in some flat directions of the supersymmetric SM, the phase-fields of the complex AD fields
have anomalous couplings to the unbroken U(1) gauge symmetry, 
and this new mechanism can be naturally realized. 
This may change the cosmological consequences of the AD mechanism, such as $Q$-ball formation~\cite{Kusenko:1997zq,Kusenko:1997si,Enqvist:1997si,Enqvist:1998en,Kasuya:1999wu,Kasuya:2000wx,Kasuya:2000sc}. 
One may wonder if our idea may spoil the AD mechanism as a mechanism for baryogenesis.  
Indeed, even if the rotating AD field generates both baryon ($B$) and lepton ($L$) asymmetries while maintaining $B$$-$$L=0$, these asymmetries are efficiently transferred to the magnetic helicity, so they become smaller. However, as shown in Ref.~\cite{Kamada:2016cnb}, the baryon asymmetry is regenerated through the transfer of the magnetic helicity during the electroweak  phase transition.  
Thus, the AD mechanism can still be responsible for the baryon asymmetry, albeit  indirectly, like the case discussed in Ref.~\cite{Kamada:2018tcs}. 

This paper is organized as follows. 
In Sec.~\ref{sec2},  we shall study the cosmological consequences of the complex scalar field with the anomalous coupling to the unbroken U(1) gauge symmetry. 
We identify its evolution, like the AD mechanism, and determine the resultant magnetic field properties generated by this new mechanism. 
Then, in Sec.~\ref{sec3} we discuss how our mechanism may be realized and embedded within well-motivated extensions of the SM. Finally,  
in Sec.~\ref{sec4}, we provide some concluding remarks and future prospects for this mechanism.

\section{magnetogenesis from a rotating scalar in the field space \label{sec2}}

\subsection{Axion-induced current as the scalar chiral magnetic effect}

First, we study a toy model as a low energy effective theory
and investigate its cosmological consequences. In the next section, we will discuss realizations of the scenario in realistic models of physics beyond the SM. Let us consider a simple model of a complex scalar field (\`a la  AD field) with an approximate global U(1)$_A$ symmetry and a massless U(1) gauge field, \begin{align}
- \frac{\cal L}{\sqrt{-g}}=&~ \partial_\mu \phi^* \partial^\mu \phi + \frac{1}{4} F_{\mu\nu} F^{\mu\nu} + (m_0^2-c_H H^2)|\phi|^2 + b_\phi (\phi^2+\mathrm{h.c.})\notag \\
& + \frac{ (a_\phi \phi^n+\mathrm{h.c.})}{n M^{n-3}}  + \frac{|\phi|^{2n-2}}{M^{2n-6}} + c_F \frac{e^2}{16 \pi^2} \theta F_{\mu\nu} {\tilde F}^{\mu\nu},  \label{toylag}
\end{align} 
motivated by the AD mechanism~\cite{Affleck:1984fy,Dine:1995kz}. 
This is enough to catch the essence of our idea.  
The scalar field $\phi$ is neutral under the U(1) gauge interaction. 
Here we adopt the metric convention $g_{\mu\nu} = (-,+,+,+)$ and 
consider the Friedmann background $ds^2 = -dt^2+a^2(t) d {\bm x}^2$ with $H={\dot a}/a$ being the Hubble parameter. 
We use the dot as the derivative with respect to the physical time $t$. 
$m_0$ is the zero-temperature mass, $c_H$ is a numerical coefficient 
of the order of the unity that parameterizes the negative Hubble induced mass, $b_\phi$ and $a_\phi$ parameterize the small global U(1)$_A$ symmetry breaking terms ($b_\phi$ and $a_\phi$-terms, respectively), and $M$ is the cutoff scale of the higher-dimensional operators. We assume that the scalar field receives the negative Hubble induced mass squared during and after inflation and the value of $c_H$ does not change significantly. 
$b_\phi$ is taken to be real while $a_\phi$ is taken to be complex without loss of generality. 
${\tilde F}^{\mu\nu} = \epsilon^{\mu\nu\rho\sigma} F_{\rho\sigma}/2\sqrt{-g}$ is the dual tensor with $\epsilon^{\mu\nu\rho\sigma}$ being the Levi-Civita symbol, $\epsilon^{0123} = 1$,  $\theta=\theta(x)$ is the phase-field of the complex scalar $\phi(x)$, $e$ is the gauge coupling constant, and $c_F$ is the numerical coefficient of the order of unity for the anomalous coupling. As the phase-field $\theta$ is the (pseudo) Nambu Goldstone boson associated with spontaneous symmetry breaking of the U(1)$_A$, it can be also regarded as an axion-like field. 

When the Hubble parameter is much larger than the zero-temperature mass, 
the net mass squared term is negative and the scalar field gets an expectation value as 
\begin{equation}
\phi = \frac{ \varphi(t) }{\sqrt{2}}e^{-i \theta(x)}. 
\end{equation}  
Once the phase of the scalar field acquires a non-zero velocity, ${\dot \theta} \not= 0$, this yields the U(1)$_A$ asymmetry in the system, 
\begin{equation} \label{nA_asymmetry}
n_A = i ({\dot \phi^*} \phi - {\dot \phi} \phi^*) =  \varphi^2{\dot \theta} . 
\end{equation}
The non-zero velocity of the angular field originates from the rotation of the complex scalar in the field space due to  the U(1)$_A$ breaking $a_\phi$-term, 
as in the case of the AD mechanism~\cite{Affleck:1984fy,Dine:1995kz}.
Taking this configuration as the background, it can easily be seen that the equations of motion for the gauge field are given by
\begin{equation}
\partial_\mu (\sqrt{-g} F^{\mu\nu}) + \sqrt{-g}c_F\frac{e^2}{4\pi^2}(\partial_\mu \theta) {\tilde F}^{\mu\nu}=0 \  \Rightarrow \  
-\frac{1}{a^2}\frac{d}{dt} (a^2 E^i)+\frac{\epsilon^{ijk}}{a(t)} \frac{\partial B_k}{ \partial x^j} - c_F\frac{e^2}{4\pi^2} {\dot \theta} B^i = 0. 
\end{equation}
Thus we determine that the current induced by the number density of the U(1)$_A$ asymmetry, 
\begin{equation}
J_{\rm ind}^i =  c_F\frac{e^2}{4 \pi^2} {\dot \theta} B^i  =  c_F\frac{e^2}{8 \pi^2} \frac{n_A(t)}{a^3(t)\varphi^2(t)} B^i ,
\end{equation}
mimics the chiral magnetic effect~\cite{Vilenkin:1980fu,Fukushima:2008xe} with a correspondence 
\begin{equation}
\mu_5 \leftrightarrow c_F\frac{ n_A}{4 a^3(t) \varphi^2}.
\end{equation} 
Here the physical electric and magnetic fields are defined as
\begin{equation}
E^i = a(t) F^{0i}, \quad E_i = a^{-1}(t) F_{i0}, \quad B_i =\frac{a^2(t)}{2} \epsilon_{i jk} F^{jk},  \quad B^i = \frac{a^{-2}(t)}{2} \epsilon^{ijk} F_{jk} .
\end{equation}
This effective current is nothing but an axion-induced current in the axion electromagnetism. 
In the literature it has been argued that the chiral magnetic effect is understood to be an effective axion field~\cite{Frohlich:2000en,Frohlich:2002fg,Boyarsky:2015faa}. 
Here we just emphasize that  by relating the axion velocity ${\dot \theta}$ to 
the number density of the U(1)$_A$ asymmetry, 
the correspondence between the chiral magnetic effect and the 
axion-induced current is clearer. 
Note that the number density of the chiral asymmetry at a high temperature $T$ 
is given in terms of the chiral chemical 
potential by $n_5 = \mu_5 T^2/6$.

\subsection{Generation of U(1)$_A$ asymmetry and magnetogenesis in the early Universe}

The axion-induced current causes tachyonic instability on the gauge fields, 
which is the essence of axionic inflationary magnetogenesis~\cite{Turner:1987bw,Garretson:1992vt,Anber:2006xt}. 
In that case, the non-zero axion velocity $\dot\theta$ is driven by a (time-independent) axion scalar potential, and the gauge field is mainly produced just after inflation, i.e. during several oscillations of the axion field~\cite{Fujita:2015iga,Adshead:2016iae,Cuissa:2018oiw}. Therefore, the corresponding current ($\propto \dot\theta$) is not a constant and even changes sign during the magnetic field amplification process. 
In this sense, the magnetic field amplification is less efficient, 
and the process is somehow different from the chiral plasma instability~\cite{Joyce:1997uy,Tashiro:2012mf,Rogachevskii:2017uyc,Brandenburg:2017rcb,Schober:2017cdw,Kamada:2018tcs}.
In contrast, if the rotation in a complex scalar field space is induced by a U(1)$_A$ breaking term that is no longer effective after its onset,  the dynamics 
of the phase-field becomes different from that of the axion mentioned above. The barrier of the scalar potential along the axion direction $\theta$ decreases over time and disappears so that the axion does not oscillate and $\dot\theta$ can be taken as a constant until the backreaction becomes important. In this case, the process is quite similar to the chiral plasma instability. In the following, we investigate the mechanism that generates the U(1)$_A$ asymmetry 
in a similar way to the AD mechanism~\cite{Affleck:1984fy,Dine:1995kz},
and the resulting magnetogenesis.

\begin{figure}[h]
	\centering
	\includegraphics[width=.60 \textwidth]{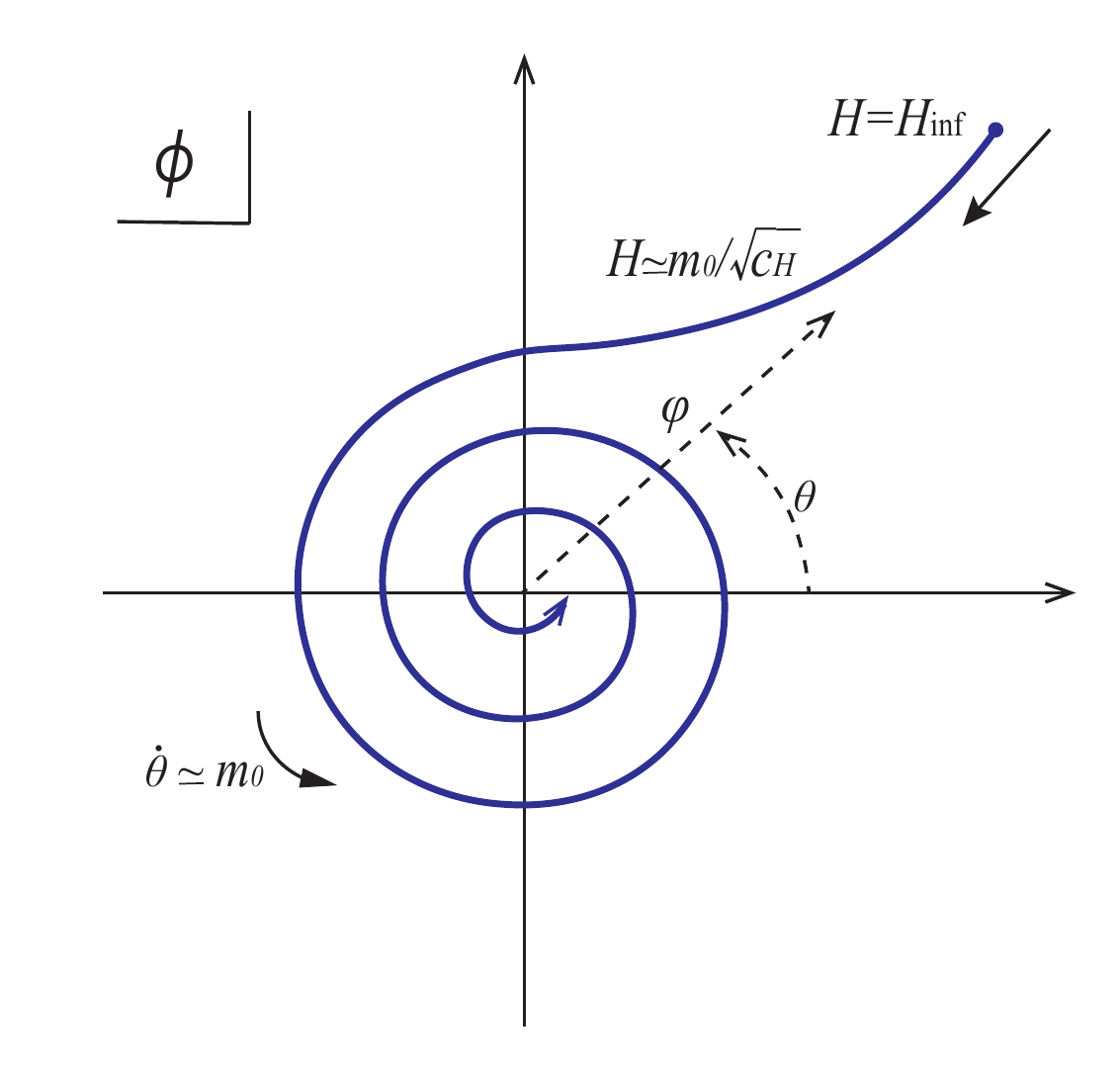}
	\caption{Schematic picture of the evolution of $\phi$ field}
	\label{fig:1}
\end{figure}

Suppose the Universe undergoes inflation, followed by a matter-dominated era due to inflaton oscillations.
Here we adopt the model with Eq.~\eqref{toylag}, assuming $m_0 \sim |a_\phi| \gg \sqrt{b_\phi}$.\footnote{If $a_\phi$ is much larger than $m_0$, unwanted symmetry-breaking vacua appear, hence we have to be more careful to ensure that $\phi$ is not trapped in the false vacua~\cite{Kawasaki:2006yb}. 
If $a_\phi$ is much smaller than $m_0$, the trajectory of $\phi$ becomes  highly elliptical with a large eccentricity so that ${\dot \theta}$ strongly  oscillates. 
In this case, magnetic fields would be generated through $\theta F_{\mu\nu} {\tilde F}^{\mu\nu}$ 
coupling but quantitative estimates get more complicated, which is beyond the scope of this study.}  
When the Hubble parameter is large
during inflation and the period of inflaton oscillation ($H>m_0/\sqrt{c_H}$), 
the $\phi$ field follows the (time-dependent) potential minimum 
generated by the balance between the negative quadratic and positive $|\phi|^{2n-6}$ term, $\varphi \simeq (H M^{n-3})^{1/(n-2)}$, 
with a spatially homogeneous distribution. 
Thanks to inflation, we  naturally suppose that the phase-field $\theta$ is also spatially homogeneous. 
	As the Hubble parameter decreases, eventually the potential minimum disappears at 
	$H_\mathrm{osc} \simeq m_0/\sqrt{c_H}$ and the $\phi$ field starts oscillation around the origin.
At the onset of oscillation, the $a_\phi$-term also gives a kick in the phase direction so that the non-zero number density of the U(1)$_A$ charge, 
\begin{equation}
n_A \simeq  \varphi_{\rm osc} ^2 {\dot \theta}, \quad \text{with} \quad \varphi \simeq  \varphi_{\rm osc} \equiv (m_0 M^{n-3})^{1/(n-2)}, \quad {\dot \theta} \simeq m_0,
\end{equation}
is generated and the trajectory of the scalar field in the complex field space is an ellipse with a small eccentricity 
for $a_\phi \sim m_0$~\cite{Dine:1995kz}.  
Here the subscript ``osc'' indicates that the quantity is evaluated at the onset of the 
scalar field oscillation. The scalar field evolve as 
\begin{equation} \label{phi_theta_motion}
\varphi \propto a^{-3/2}, \quad {\dot \theta} \simeq m_0 = \text{const}. 
\end{equation}
The former in Eq.~(\ref{phi_theta_motion}) comes from the fact that both the real and imaginary parts of the scalar field are harmonic oscillators in the matter-dominated Universe and damp in proportion to $t^{-1}$,  and the latter is derived from the comoving number density conservation, $a^3 n_A = a^3 {\dot \theta} \varphi^2 = \text{const}$.  During the evolution, the U(1)$_A$-breaking $a_\phi$-term potential decays in proportion to $a^{-3n/2}$ whereas the quadratic term scales as $a^{-3}$. 
Hence, roughly in Hubble time after the onset of oscillation,  the U(1)$_A$ breaking term becomes ineffective and $n_A$ is nearly preserved as long as the $b_\phi$-term is negligible.  
Figure \ref{fig:1} shows the schematic picture of the evolution of the $\phi$ field. 

Before proceeding, let us comment about an issue omitted in the discussion above. 
Indeed,  to show our idea simply and clearly, we do not take thermal effects  into account~\cite{Allahverdi:2000zd,Anisimov:2000wx}. 
In principle, there should be thermal corrections to the scalar potential 
even before the completion of reheating, since 
the partial decay of inflaton quanta generates a high temperature plasma as a subdominant
component of the Universe. The absence of such thermal corrections are valid if, {\it e.g.}, the 
inflaton decays mainly into a hidden sector and the SM particles are not significantly
produced. 
If thermal corrections to the scalar field potential exist, 
they induce an early onset of the scalar field oscillation, 
whose eccentricity is larger.

Concretely, this requirement is satisfied if
the thermal correction is smaller than the bare mass term at the onset of $\phi$  oscillation. 
Typically the thermal potential  is given as~\cite{Allahverdi:2000zd,Anisimov:2000wx}
\begin{equation}
V_\mathrm{th}(\varphi) \simeq \left\{\begin{array}{ll} \dfrac{1}{2} T_\mathrm{SM}^2 \varphi^2 & \quad \text{for} \ T>  \varphi, \\ 
T_\mathrm{SM}^4 \log \left(\dfrac{\varphi^2}{T^2}\right) &\quad \text{for} \ T< \varphi, 
\end{array} \right. 
\end{equation} where  $T_\mathrm{SM}$ is the temperature of the Standard Model plasma. We did not explicitly write the coupling constants of order of unity that give the dominant contribution for the 
thermal potential (typically gauge couplings or the top Yukawa coupling).  The upper one is the thermal mass which is active when the fields directly coupled to the $\phi$ field are light enough to be in thermal equilibrium. If the masses of the coupled fields are heavier than the temperature, $M_\mathrm{f} \simeq \varphi > T_\mathrm{SM}$, as will be discussed in  Sec.~\ref{2HDM}, 
a two-loop contribution gives the lower one, the thermal logarithmic potential~\cite{Anisimov:2000wx}.
If we suppose that the inflaton decay rate is much smaller than the inflaton mass, the reheating process is well described by the perturbative decays of the inflaton field.
The temperature of the hidden sector $T_\mathrm{hid}$ as well as that of the  
Standard Model sector before the completion of reheating are given by~\cite{kolbturner}
\begin{equation} \label{reheating}
T_\mathrm{hid} \simeq (M_\mathrm{pl} ^2H \Gamma_\mathrm{hid})^{1/4}, \quad T_\mathrm{SM} \simeq (M_\mathrm{pl}^2 H \Gamma_\mathrm{SM})^{1/4}, 
\end{equation}
where $M_\mathrm{pl}$ is the reduced Planck mass and 
$\Gamma_\mathrm{hid}$ ($\Gamma_\mathrm{SM}$)
is the decay rate of the inflaton into the hidden sector (the Standard Model sector).
Note that the former of Eq.~(\ref{reheating}) is related to the reheating temperature of the hidden sector  as 
$ \Gamma_\mathrm{hid} \simeq T_\mathrm{RH}^2/M_\mathrm{pl}$.
Since $T_{\rm SM}$ decreases more slowly than the Hubble parameter, 
it is important to evaluate the thermal potential at around $H\simeq m_0$. 
For $m_0 \ll \varphi_\mathrm{osc}$ which is the case of our interest, the thermal logarithmic potential must be smaller 
than the Hubble induced potential. Thus, $T_\mathrm{SM}\ll \sqrt{\varphi_\mathrm{osc}  m_0}$ and
\begin{equation} \label{constgsm2}
\Gamma_{\rm SM} \ll 2\times   10^{-18}{\rm GeV} \left(\frac{m_0}{10^3{\rm GeV}}\right)\left(\frac{\varphi_\mathrm{osc}}{10^8{\rm GeV}}\right)^2.
\end{equation} 
In terms of  the branching ratio, the constraint reads 
\begin{equation} \label{constbr2}
{\rm Br}_{\rm SM} \ll  4\times  10^{-16} 
\left(\frac{m_0}{10^3{\rm GeV}}\right)\left(\frac{\varphi_\mathrm{osc}}{10^8{\rm GeV}}\right)^2
\left(\frac{T_{\rm RH}}{10^8\,{\rm GeV}}\right)^{-2}.
\end{equation} 
The small branching ratio might be achieved by tiny couplings ($g_{\rm SM} \ll g_{\rm hidden}$) or by kinematics ($M_{\rm hidden}\ll  M_{\rm inflaton}\ll M_\mathrm{f}\sim \varphi_{\rm osc}  $) during magnetogenesis.  A detailed implementation of viable reheating models is an interesting problem, but beyond the scope of the paper, so we leave it as future work.

Now let us examine how the gauge fields are amplified due to the tachyonic instability 
and how they backreact to the scalar field dynamics. The equations of motion for the 
phase direction of the scalar field and gauge fields are given by 
\begin{align}
&\partial_\mu (\sqrt{-g} \varphi^2(t) \partial^\mu \theta) - \sqrt{-g}c_F \frac{e^2}{16 \pi^2} F_{\mu\nu} {\tilde F}^{\mu\nu} = 0, \label{cons} \\
& \partial_\mu\left( \sqrt{-g}F^{\mu\nu} \right)+ \sqrt{-g} c_F\frac{e^2}{4\pi^2} (\partial_\mu \theta){\tilde F}^{\mu\nu} =0. 
\end{align}
The latter exhibits the instability of the gauge fields for the non-zero background ${\dot \theta}$.  It can be explicitly seen as follows. 
As long as the phase-field evolves with a homogeneous constant velocity, 
$\partial_\mu \theta \simeq ({\dot \theta}, 0,0,0)$, with a negligible backreaction, 
we can take them as a background for the evolution of the gauge fields. 
Switching from the physical time to the conformal time so that  
$ds^2 = a^2(\tau) ( - d \tau^2+d {\bm x}^2)$, the equations of motion for the gauge fields read
\begin{equation}
- \frac{\partial^2}{\partial \tau^2} A_i+ \sum_j \frac{\partial^2}{\partial x_j^2} A_i+ c_F \frac{e^2 a(\tau)}{4 \pi^2}\dot \theta  \sum_{j,k} \epsilon_{ijk}  \partial_j A_k =0, 
\end{equation}
where we work in the radiation gauge ${\bm \nabla} \cdot {\bm A} =0, A_0=0$. 
To solve the equations of motion, it is convenient to work in the momentum space 
by performing a Fourier transformation, 
\begin{equation}
A_i(\tau, {\bm x}) = \sum_{\lambda=\pm} \int \frac{d^3 k}{(2\pi)^{3/2}} \left[A_\lambda(\tau, {\bm k}) \epsilon_{i,\lambda}({\bm k}) e^{i {\bm k} {\bm x}}  +{\rm h.c.} \right] ,
\end{equation}
with $\epsilon_{i,h}({\bm k})$  being the circular polarization tensor that satisfies 
\begin{equation}
\epsilon_{i,\lambda} ({\bm k}) \cdot \epsilon^{*i}_{\lambda'}({\bm k}) = \delta_{\lambda,\lambda'} \quad k^i \epsilon_{i,\lambda} ({\bm k}) = 0, \quad i \epsilon^{ijk} k_j \epsilon_{k,\lambda} ({\bm k}) = \lambda k \epsilon_{i,\lambda} ({\bm k}). 
\end{equation}
With these decomposition the equations of motion for the Fourier modes are rewritten as
\begin{equation}
- \frac{\partial^2}{\partial \tau^2} A_\lambda(\tau, {\bm k}) - k^2 A_\lambda(\tau, {\bm k}) +  c_F\frac{e^2 a(\tau)}{4\pi^2} {\dot \theta} \lambda k A_\lambda(\tau, {\bm k}) =0.  \label{eomgf}
\end{equation}
We see that the last term acts as a tachyonic mass term for ${\dot \theta} \lambda >0$ ($\lambda = \pm 1$) and triggers the instability of the gauge fields. 
For the inflaton oscillation epoch with $a(t) \propto t^{2/3}\propto \tau^2$, 
the $\pm$ mode of the gauge field feels unstable for ${\dot \theta} \gtrless 0$ 
at $k_\mathrm{ins}/a(\tau_\mathrm{ins}) \simeq c_F e^2 {\dot \theta}/4 \pi^2$ around $a(\tau_\mathrm{ins}) \tau_\mathrm{ins} \simeq 4\pi^2 /(c_Fe^2 {\dot \theta})$, equivalently $H_\mathrm{ins} \simeq c_F e^2 {\dot \theta}/8\pi^2 \simeq c_F e^2 m_0/8\pi^2$.  
As a result, for a given sign of $\dot\theta$ just one mode grows exponentially, 
so  maximally helical gauge fields are obtained.
Here the subscript ``ins'' indicates that the quantity is evaluated at the time 
when the instability starts to grow. 
Here we assume that there is no thermal plasma in the Standard Model sector, which includes relevant U(1) gauge charged particles, as has also been discussed in the $\phi$ field dynamics. 

It also modifies the equations of motion of the gauge fields \eqref{eomgf}, by introducing a friction term $- \sigma_\mathrm{SM} \partial A_\lambda/\partial \tau$ with $\sigma_\mathrm{SM} \simeq 100 T$~\cite{Baym:1997gq,Arnold:2000dr} being the electric conductivity
induced by the SM plasma. 
 Light charged degrees of freedom would also induce 
electric currents like the Schwinger effect. 
Then the magnetic field amplification would become less efficient~\cite{Domcke:2018eki} 
and  the light particles may be thermalized~\cite{Tangarife:2017rgl}. 
As a result, the process of gauge field amplification becomes more involved. 
In light of this, here we assume that there are no light charged degrees of freedom, 
which can be satisfied  if all  charged degrees of freedom acquire their masses from the AD field.
Investigation of this effect is left for a future study.  
See also the discussion at the end of Sec.~\ref{sec:3B}.

The amplification of the gauge fields stops when the backreaction  
 {from the gauge field production becomes non-negligible. 
By taking a spatial average, Eq.~\eqref{cons} can be understood as the conservation law for the sum of chiral asymmetry and magnetic helicity, 
\begin{equation}
\frac{\partial}{\partial \tau} \left(a^3 (\tau)\varphi^2 {\dot \theta} +c_F\frac{e^2}{8 \pi^2}h \right) =0, \quad h = \frac{1}{V}\int_V d^3 x  \epsilon^{ijk} A_i \partial_j A_k . 
\end{equation}
Therefore we can estimate that when 
\begin{align}
&c_F\frac{e^2}{8 \pi^2} h_\mathrm{sat} = \frac{c_F}{V}\frac{e^2}{8 \pi^2} \int_V d^3 x  \epsilon^{ijk} A_i \partial_j A_k  \simeq a^3(\tau_\mathrm{ins}) \varphi^2(\tau_\mathrm{ins}) {\dot \theta} \simeq a^3(\tau_\mathrm{osc})  \varphi^2(\tau_\mathrm{osc}) {\dot \theta} \notag \\
&\Leftrightarrow h_\mathrm{sat} = \frac{8 \pi^2}{c_F e^2}a^3(\tau_\mathrm{osc})  \varphi^2_\mathrm{osc} {\dot \theta}
\end{align}
the amplification of the gauge fields saturates. 
In other words, magnetic field amplification stops when the maximal transfer from chiral asymmetry to magnetic helicity is completed. 
Here the subscript ``sat'' indicates that the quantity is evaluated at the time when  
the gauge field amplification becomes saturated.  Since the instability induces an exponential growth, 
we approximate $\tau_\mathrm{sat} \simeq \tau_\mathrm{ins}$. 
Focusing on the magnetic fields, by approximating 
\begin{equation}
h_\mathrm{sat} \simeq  a^2(\tau_\mathrm{sat})  A_\mathrm{sat}  B_\mathrm{sat} =  a^3(\tau_\mathrm{ins})\frac{B_\mathrm{sat}^2}{k_\mathrm{ins}/a(\tau_\mathrm{ins})},  \quad \text{where} \quad B_\mathrm{sat} = \frac{k_\mathrm{ins}}{a^2(\tau_\mathrm{ins})} A_\mathrm{sat}, 
\end{equation}
the magnetic field strength $B$ and coherence length $\lambda_B$ 
are obtained 
\begin{align}
B_\mathrm{sat}  \simeq &~ \sqrt{\frac{8\pi^2}{c_F e^2}} \left(\frac{a(\tau_\mathrm{osc})}{a(\tau_\mathrm{ins})}\right)^{3/2}  \varphi_\mathrm{osc} \sqrt{ \frac{k_\mathrm{ins}{\dot \theta}}{a(\tau_\mathrm{ins})}}  \simeq \sqrt{2} \left(\frac{a(\tau_\mathrm{osc})}{a(\tau_\mathrm{ins})}\right)^{3/2} \varphi_\mathrm{osc} {\dot \theta} \notag \\
\simeq&~ \sqrt{2}  \left(\frac{H_\mathrm{ins}}{H_\mathrm{osc}}\right) \varphi_\mathrm{osc} {\dot \theta}
\simeq 2 \times 10^{12}\, \mathrm{GeV}^2  \left(\frac{\varphi_\mathrm{osc}}{ 10^{12}\, \mathrm{GeV}}\right)\left(\frac{\dot \theta}{10^3\, \mathrm{GeV}}\right), \\ 
\lambda_{B,\mathrm{sat}} \simeq &~ \lambda_B (\tau_\mathrm{ins}) \simeq 2 \pi \left(\frac{k_\mathrm{sat}}{a(\tau_\mathrm{sat})}\right)^{-1}  \simeq 2 \pi \left(c_F \frac{e^2}{4 \pi^2} {\dot \theta}\right)^{-1} \simeq c_F^{-1} \left(\frac{3}{1 \mathrm{GeV}}\right) \left(\frac{ \dot \theta}{10^3\, \mathrm{GeV}}\right)^{-1}
\end{align}
at the time when the gauge field amplification gets saturated.
Here we take $e\simeq 0.3$. 
It is noted that $B_\mathrm{sat}$ is independent of $c_F$ while $\lambda_{B,\mathrm{sat}}$ is inversely proportional to $c_F$. Note also that in the absence of thermal plasma, electric fields in amounts similar to the magnetic 
fields are produced at the same time. 

\subsection{Cosmological evolution of magnetic fields}

So far we have not specified the relationship between the
U(1) gauge symmetry discussed in the previous sections and the U(1) in the SM.  
Let us investigate the cosmological consequences when the gauge fields are those of the U(1) gauge symmetry in the SM.  After the saturation of the gauge field amplification,  the physical magnetic field (as well as the electric field) evolves adiabatically,  $B \propto a^{-2}$ and $\lambda_B \propto a$,  
until the SM particles are thermalized and the magnetohydrodynamics becomes important 
for their evolution~\cite{Fujita:2016igl,Jimenez:2017cdr}. 
Once the SM particles are thermalized, the electric fields are screened due to the thermal effect,
while the magnetic fields retain their properties. 
The magnetic fields induce the fluid dynamics and the fluid develops a turbulence.  
Then both the magnetic fields and velocity fields start to co-evolve according to the magnetohydrodynamic equations and follow the inverse cascade process once the eddy turnover scale of the fluid catches up with the magnetic field coherence length, $\lambda_B \simeq v_A t \simeq B/\sqrt{\rho} H$,  where $v_A$ is the Alfv\'en velocity~\cite{Banerjee:2004df,Kahniashvili:2012uj}. 
The magnetic field further evolves until today according to the magnetohydrodynamics, 
which determines the linear relation between
the magnetic field strength and coherence length today as~\cite{Banerjee:2004df}
\begin{equation}
\lambda_B (t_0) \sim  1 \mathrm{pc} \left( \frac{B(t_0)}{10^{-14}\mathrm{G}}\right),  \label{blamtoday}
\end{equation}
where $t_0$ is the present physical time.  On the other hand, thermal plasma induces a large electric conductivity, which ensures the comoving magnetic helicity is a good conserved quantity. 
Since it is also conserved during adiabatic evolution, we have the relation 
\begin{equation}
a(t_0)^3 \lambda_B(t_0) B(t_0)^2 \simeq a(\tau_\mathrm{ins})^3 \lambda_B(\tau_\mathrm{ins}) B(\tau_\mathrm{ins})^2. 
\end{equation}
Then we have 
\begin{align}
\lambda_B(t_0) B(t_0)^2 =&~ \left(\frac{a_\mathrm{RH}}{a(t_0)}\right)^3\left(\frac{a_\mathrm{ins}}{a_\mathrm{RH}}\right)^3  \lambda_\mathrm{ins} B_\mathrm{ins}^2 \notag \\
=&~ \frac{g_{*s}^0 T_0^3}{g_{*s}^\mathrm{RH} T_\mathrm{RH}^3} \left(\frac{H_\mathrm{RH}}{H_\mathrm{ins}}\right)^2 \left(\frac{12\times 10^{24} \mathrm{GeV}^3 }{c_F}\right) \left(\frac{\varphi_\mathrm{osc}}{10^{12}\, \mathrm{GeV}}\right)^2 \left(\frac{\dot \theta}{10^3\, \mathrm{GeV}}\right) \notag \\
=&~ 9 \times 10^{-68} \left(\frac{T_\mathrm{RH}}{10^8\,\mathrm{GeV}}\right) \left(\frac{H_\mathrm{ins}}{\mathrm{GeV}}\right)^{-2} \left(\frac{12\times 10^{24} \mathrm{GeV}^3 }{c_F}\right)
 \left(\frac{\varphi_\mathrm{osc}}{10^{12} \mathrm{GeV}}\right)^2
 \left(\frac{\dot \theta}{10^3\, \mathrm{GeV}}\right) \notag \\
=&~\left(\frac{10^{-35}\, \mathrm{pc} \mathrm{G}^2}{c_F}\right)    \left(\frac{T_\mathrm{RH}}{10^8\,\mathrm{GeV}}\right) \left(\frac{H_\mathrm{ins}}{ \mathrm{GeV}}\right)^{-2} \left(\frac{\varphi_\mathrm{osc}}{10^{12}\, \mathrm{GeV}}\right)^2\left(\frac{\dot \theta}{10^3 \mathrm{GeV}}\right), 
\end{align} 
where we have used $g_{*s}=3.91, T_0 = 2.3 \times 10^{-13}\, \mathrm{GeV}, 1 \mathrm{pc} = 1.56 \times 10^{32}\, \mathrm{GeV}^{-1}$, and $1\, \mathrm{G} = 1.95 \times 10^{-20}\, \mathrm{GeV}^2$ 
(in natural Lorentz-Heaviside units)
and assumed that at $H = H_\mathrm{RH}$, the Universe is filled with relativistic particles with the effective temperature $T_\mathrm{RH}$, and 
the energy density and entropy are given by $\rho=(\pi^2 g_{*s}^\mathrm{RH}/30) T_\mathrm{RH}^4, 
s = (2 \pi^2 g_{*s}^\mathrm{RH}/45)T_\mathrm{RH}^3$. 
We have also assumed that  the Universe is eventually filled with the SM radiation without additional entropy production. 
Combining it with Eq.~\eqref{blamtoday}, and assuming $c_F\simeq 1$, we obtain the present magnetic field properties, 
\begin{align}
&B(t_0) \simeq 10^{-16}\, \mathrm{G}\left(\frac{T_\mathrm{RH}}{10^8\,\mathrm{GeV}}\right)^{1/3} \left(\frac{H_\mathrm{ins}}{ \mathrm{GeV}}\right)^{-2/3}\left(\frac{\dot \theta}{10^3\, \mathrm{GeV}}\right)^{1/3} \left(\frac{\varphi_\mathrm{osc}}{10^{12}\, \mathrm{GeV}}\right)^{2/3}, \\ 
&\lambda_B(t_0)\simeq 10^{-2}\, \mathrm{pc} \left(\frac{T_\mathrm{RH}}{10^8\,\mathrm{GeV}}\right)^{1/3} \left(\frac{H_\mathrm{ins}}{\mathrm{GeV}}\right)^{-2/3}\left(\frac{\dot \theta}{10^3\, \mathrm{GeV}}\right)^{1/3} \left(\frac{\varphi_\mathrm{osc}}{10^{12}\, \mathrm{GeV}}\right)^{2/3}. 
\end{align}
This suggests that the detection of intergalactic magnetic fields with maximal helicity 
can be a trace of this scenario. 

Moreover, we note that 
the set of  fiducial values is suitable for baryogenesis~\cite{Kamada:2016cnb}. 
This is not surprising because if there is not a magnetic field amplification and the asymmetry is 
conserved, the asymmetry-to-entropy ratio is 
\begin{equation}
\frac{n}{s} = \left(\frac{a_\mathrm{osc}}{a_\mathrm{RH}}\right)^3 \frac{{\dot \theta} \varphi_\mathrm{osc}^2}{(2 \pi^2 g_{*s}^\mathrm{RH}/45)T_\mathrm{RH}^3} \sim 10^{-9}, 
\end{equation}
for the fiducial values. 
In this scenario, if the generated magnetic fields are those of hypergauge interaction, 
the asymmetry produced by the scalar field dynamics is first transferred to  
hypermagnetic helicity. It is eventually transferred back to the baryon asymmetry 
at the electroweak phase transition, without large loss in the sum of magnetic helicity and 
U(1)$_A$ asymmetry, similar to the case studied in Ref.~\cite{Kamada:2018tcs}. 
Even if electroweak symmetry is broken down to the electromagnetism by the expectation values of the scalar field and the electromagnetic fields are produced in this scenario, they transform into the hypermagnetic fields once the scalar field decays. Then the same process follows for baryogenesis.

\subsection{Comment on the $b_\phi$-term \label{sec:2d}} 

The effect of the $b_\phi$-term has been ignored to avoid the time variation
in the U(1)$_A$ asymmetry. However, from the phenomenological point of view, this term is unavoidable in some realizations. We discuss how small this term should be for successful magnetogenesis.

Let us examine the evolution of the scalar fields in more depth after the onset of oscillation. 
Taking into account the $b_\phi$-term, the masses of the real and imaginary parts of 
the complex scalar field differ  as 
\begin{equation}
m_\mathrm{re/im} = \sqrt{m_0^2 \pm 2 b_\phi} \equiv m_0 \pm \Delta m. 
\end{equation}
When $b_\phi$ is hierarchically smaller than $m_0^2$, $\Delta m\simeq b_\phi/m_0 \ll m_0$. 
The evolution of the scalar fields is given by 
\begin{align}
\phi_R(t) \equiv&~ \mathrm{Re} (\phi(t)) \simeq  
 \sqrt{m_0 M}  \left(\frac{\cos (m_\mathrm{re}  t)}{m_0t}\right), \nonumber\\
\phi_I(t)\equiv&~ \mathrm{Im}(\phi(t)) \simeq   \sqrt{m_0 M}   \left(\frac{\sin (m_\mathrm{im}  t)}{m_0 t} \right), 
\end{align}
which yields
\begin{align}
{\dot \theta} =&~ \frac{ \phi_R(t) \dot\phi_I(t) -  \dot\phi_R(t)  \phi_I(t)}{ \phi_R^2(t) + \phi_I^2(t) }  
\simeq \frac{m_0 \cos(2 \Delta m t)  - \Delta m \cos (2 m_0 t)}{1 - \sin (2m_0 t) \sin (2 \Delta m t)}. \label{dotthetab}
\end{align}
${\dot \theta}$ evolves with the combination of the oscillation with 
a longer period $\Delta t_L \simeq (\Delta m)^{-1}$ and 
the one with a shorter period $\Delta t_S\simeq (m_0)^{-1}$. 
This means that after the onset of oscillation, the trajectory of the scalar in the field space 
is approximately a circle, as long as  $t\ll \Delta t_L$, so that we can take $\dot\theta$ as a constant. Let us adopt an ansatz that ${\dot \theta}$ is regarded as a constant if  $0.9 m_0 \lesssim {\dot \theta} \lesssim 1.1 m_0$.  
This corresponds to $\sin (\Delta m t) \lesssim 0.1$. Thus we take 
\begin{equation}
\Delta t_\mathrm{c} = 0.1 (\Delta m)^{-1} 
\end{equation}
as the criteria for the duration during which ${\dot \theta}$ can be regarded as a constant.  
For $t \gtrsim \Delta t_c$, eventually it becomes decoherent and ${\dot \theta}$ 
cannot be taken as a constant any longer.

Since the magnetic field amplification occurs 
within the time scale
\begin{equation}
\Delta t  \sim H_\mathrm{ins}^{-1} \simeq \left(c_F\frac{e^2}{16 \pi^2} {\dot \theta}\right)^{-1}\simeq \left(c_F\frac{e^2}{16 \pi^2} m_0 \right)^{-1}, 
\end{equation}
requiring that this is  shorter than $\Delta t_\mathrm{c}$, 
we obtain the constraint on $\Delta m$ as
\begin{equation}
\Delta m < c_F\frac{e^2}{160 \pi^2} m_0 \quad \text{or} \quad b_\phi < c_F \frac{e^2}{160 \pi^2} m_0^2 \simeq 5 \times 10^{-5} c_F m_0^2. \label{bcons}
\end{equation}
This gives a constraint on the $b_\phi$-term in the 
phenomenological model building during magnetogenesis.

\section{realization \label{sec3}}

In this section, we describe how the low energy effective Lagrangian Eq~\eqref{toylag} is realized in the well-motivated models.  The idea is completely analogous to the couplings between axions and gauge fields. Namely, for large values of $\varphi \gg m_0\sim a_\phi ={\cal O}(0.1-1\,{\rm TeV})$, 
U(1)  charged fields get heavy, $M_\mathrm{f}\sim y_f\varphi$, from the interactions like 
$y_f \phi\bar  \psi_L \psi_R + {\rm h.c.}$. 
After integrating out those heavy fermions, an anomalous coupling in the form $(e^2 \theta/16 \pi^2) F_{\mu\nu} {\tilde F}^{\mu\nu}$ is induced by the triangle diagrams.  
Note that the relevant light degrees of freedom are $\theta$, and U(1) gauge field, $A_\mu$. 
We will demonstrate several examples in well-motivated models of the physics beyond the SM as proofs of concept. This suggests that such an anomalous coupling and magnetogenesis are general features 
of the AD mechanism and other similar cosmological scenarios.

\subsection{Two Higgs Doublet Model} \label{2HDM}

The first (clear) example is that the phase-field $\theta$ is the angular direction of the Higgs field in the type-II two Higgs doublet model (2HDM). 
Since it is nothing but the PQWW axion or the CP-odd Higgs field, by mapping the global U(1)$_A$ to the approximate Peccei-Quinn symmetry 
U(1)$_\mathrm{PQ}$ ($H_1  H_2 \to e^{-i\beta} H_1   H_2 $), 
we obtain the unbroken gauge symmetry U(1)$_\mathrm{em}$ for the large Higgs expectation values.  In this case, all U(1)$_\mathrm{em}$ charged SM fermions 
and vector bosons get heavy, and 
the anomalous coupling between the light CP-odd Higgs and the U(1)$_\mathrm{em}$ gauge field is generated at low energies.  Therefore, we expect the effective Lagrangian in the form of Eq.~\eqref{toylag}.  Let us see in more depth how to realize our situation of interest in the type-II 2HDM, and especially, how to realize the coherent motion of the Higgs fields and the vanishingly small $b$-term as discussed in Sec.~\ref{sec:2d}.

\subsubsection{Scalar potential}
Let us first investigate how to construct the scalar potential that allows the Higgs fields to develop large expectation values during inflation  
The SM gauge charges and PQ charges for the SM fields in the type-II 2HDM are given in Table~\ref{charge1},
\begin{table}[h]
 \begin{center}
\begin{tabular}{|c||c|c|c|c|c|c|c|c|}
\hline
 Fields& \, $Q_{Li}$\,   & \, $ u_{R i}$ \, & \, $ d_{R i}$\,  &\,  $ L_{L i} $\,  &\, $ e_{Ri}$\, &\,  $H_1 $\,  &\, $H_2$\, \\
\hline
SU(2)$_L$ & ${\bf 2}$ & ${\bf 1}$ & ${\bf 1}$& ${\bf 2}$   &  ${\bf 1}$& ${\bf 2}$ & $ {\bf 2}$\\
\hline
U(1)$_Y$ & $1/6$ & $2/3$ & $-1/3$& $-1/2$   &  $-1$& $1/2$ & $ -1/2$\\
\hline
U(1)$_{PQ}$ & $1$ & $-1$ & $-1$& $ 1$   &  $-1$& $2$ & $ 2$\\
 \hline
\end{tabular}
\end{center} 
\caption{The SU(2)$_L$$\times$ U(1)$_Y$ and PQ charge assignment in the SM.}
\label{charge1}
\end{table}
which allow us to determine  the Yukawa couplings as
\begin{equation}  -\frac{{\cal L}_{\rm Yuk}}{\sqrt{-g}} =  (y_d)_{ij} \bar Q_{Li} d_{Rj} H_1+   (y_u)_{ij} \bar Q_{Li} u_{R j} H_2 +  (y_e)_{ij} \bar L_{L i} e_{R j} H_1 + \mathrm{h.c}..  
\end{equation}
For the Lagrangian of the Higgs sector
\begin{equation}
-\frac{{\cal L}_{\rm Higgs}}{\sqrt{-g}}  = |D_\mu H_1|^2 + |D_\mu H_2|^2  + V(H_1, H_2) ,
\end{equation} 
the form of the scalar potential $V(H_1, H_2)$ is important to realize our setup. 
Note that the PQ symmetry is anomalous under the hypergauge interaction.

There are eight degrees of freedom of the Higgs fields in total, which are characterized in terms of the four complex scalars as 
\begin{equation}
H_1 = \left(\begin{array}{c} H_1^+ \\ H_1^0 \end{array}\right), \ 
H_2 = \left(\begin{array}{c} H_2^0\\ H_2^-\end{array}\right). 
\end{equation}
Indeed, we can construct scalar potentials with a flat direction using a complex scalar degree of freedom among the four, while the other six degrees of freedom are heavy enough along the flat direction. To realize such a feature,  some key ideas can be borrowed from the supersymmetric (SUSY) extension of the Standard Model for illustration.  Two Higgs doubles  are naturally introduced, and  there are three contributions to the Higgs potential, namely, from the $D$-term, $F$-term, and soft breaking terms.  Assuming that the PQ symmetry of the Higgs sector is broken only by the following higher dimensional superpotential,  
\begin{equation}\label{super}
W_\mathrm{PQB} = \frac{(H_1 H_2)^2}{M}, 
\end{equation}  
the scalar potential of $H_1$ and $H_2$ is obtained as 
\begin{align} \label{V_AD}
V_{\rm AD}(H_1, H_2) =&~ m_1^2 |H_1|^2 + m_2^2 |H_2|^2   + \left( \frac{a_H}{M} (H_1 H_2)^2 + \mathrm{h.c.} \right) + \frac{4(|H_1|^2+ |H_2|^2)|H_1  H_2|^2}{M^2} \nonumber\\
  &+  \frac{g^2+ g'^2}{8} \left(|H_1|^2 - |H_2|^2\right)^2 
+ \frac{g^2}{2} \left| H_1^+ H_2^{0*} + H_1^0 H_2^{-*}\right|^2. 
\end{align}
Here  $H_1H_2\equiv \epsilon_{a b} H_1^a H_2^b$ and $M$ is the cutoff scale. 
The first three terms in the RHS in Eq.~(\ref{V_AD}) are the soft SUSY breaking terms with $m_1 \sim m_2 \sim a_H={\cal O}(0.1-1{\rm TeV})$.  The last term in the first line is the $F$-term contribution. 
The quartic potential in the second line is the $D$-term potential, which gives the approximate flat direction:
\begin{equation} 
(H_1)_{\textrm{$D$-flat}} = \left(\begin{array}{c}  0 \\ \frac{1}{2} \varphi e^{-i\theta} \end{array}\right), \ 
(H_2)_{\textrm{$D$-flat}} = \left(\begin{array}{c}  \frac{1}{2} \varphi e^{-i\theta} \\  0\end{array}\right). 
\end{equation}
Note that once the Higgs fields develop the expectation values along the flat direction, 
the coupled charged Higgs get heavy and their expectation values vanish, 
and hence there is no $F$-term (and $D$-term) contribution from them.
Focusing on the flat direction, parameterized by the fields $\varphi$ and $\theta$, 
we obtain the effective potential of the \`a la AD field ($\varphi$ and $\theta$) in the form of Eq.~\eqref{toylag} 
(without the Hubble induced mass).

Let us check whether the other six degrees of freedom become sufficiently heavy along the flat direction. Along this direction,  taking $\varphi \gg m_0$, 
we can see the splitting of the mass spectrum into heavy modes with masses of ${\cal O}(\varphi)$, and light modes as follows. 
As ${\rm SU(2)}_L\times {\rm U(1)}_Y$ is spontaneously broken to
U(1)$_\mathrm{em}$, and denoting the fields along the flat direction as $\delta H$, three scalar degrees,  $G^0 \equiv {\rm Im}(\delta H_1^0 - \delta H_2^0)$
and $G^+\equiv \delta H_1^+ - \delta H_2^{-*}$, 
$G^- \equiv G^{+*}$ 
are eaten by $Z^0$ and $W^\pm$ and become heavy with masses $g\varphi/2$ and $\sqrt{g^2+ g'^2}\varphi/2$, respectively. One of the CP-even Higgs degrees of freedom $H^0\equiv {\rm Re}(\delta H_1^0 - \delta H_2^0)$ and the charged Higgs components $H^+\equiv\delta H_1^+ + \delta H_2^{-*}$, $H^- \equiv H^{+*}$ are also heavy with masses $
\sqrt{g^2 + g'^2} \varphi/2$ and $ g \varphi/2$ at the leading order.  
The scalar fields $\varphi$ and $\theta$ get masses only from soft terms and a higher dimensional operator, so they are much lighter than the above six scalar degrees of freedom. It is clear that the $\theta$ field is the CP-odd Higgs/the PQWW axion.

The negative Hubble induced mass terms for $H_1$ and $H_2$ can be added as
\begin{equation}
\Delta V_{\rm Hubble}= -c_1 H^2 |H_1|^2 - c_2 H^2 |H_2|^2, 
\end{equation}
by supposing, {\it e.g.}, the non-minimal couplings to gravity, $-\xi_1 R |H_1|^2-\xi_2 R |H_2|^2$ with $R$ being the Ricci scalar, 
or non-trivial K\"ohler potential between the inflaton and the Higgs 
doublets in the supersymmetric case~\cite{Dine:1995kz,Kasuya:2006wf}. 
Note that the Ricci scalar is $R={\cal O}(H^2)$ during inflation and a matter-dominated Universe.

In Eq.~\eqref{V_AD}, we did not address the $b_{H}$-term potential presented in Eq.~\eqref{toylag} (i.e. $\Delta V  = b_H H_1 H_2 + \mathrm{h.c.}$). 
A sizable $b_H$-term  is diadvantageous for generating magnetic fields as discussed in Sec.~\ref{sec:2d}. 
However, if $b_H$ is  much smaller than $m_1^2+ m_2^2$ as required, 
the value of $\langle|H_2^0|\rangle$ in the present Universe is too small to be 
realistic, because
$\langle |H_2^0|\rangle/\langle |H_1^0|\rangle \simeq |b_H|/(m_1+ m_2)^2$. This leads to non-perturbatively large Yukawa couplings to obtain the correct masses of down quarks and charged leptons, $m_{d/e}=y_{d/e}\langle |H_2^0|\rangle$.
One way to avoid this problem and give more freedom to the $b_H$-term is to consider 
the case where the $b_H$-term in the present Universe is dominated by the vacuum expectation value of a scalar field as
$b_H\sim \langle S^2\rangle={\cal O}(m_1^2+ m_2^2) $, by introducing a gauge singlet PQ charged complex scalar field, $S$, 
while the PQ breaking bare $b_H$-term is vanishingly small. Let us consider the following potential for the $S$ field, 
\begin{equation}
\Delta V_{b\textrm{-term}} = (m_S^2 +\kappa_1  |H_1|^2 + \kappa_2 |H_2|^2)|S|^2 
+( \kappa H_1 H_2 S^2 +   a_S^3 S + \mathrm{h.c.}) + \frac{\lambda_S}{4}|S|^4. 
\end{equation} 
Here $|m_S|\sim |a_S|  = {\cal O}(m_1^2+ m_2^2)$,  $\kappa, \kappa_1, \kappa_2$, and $\lambda_S$ are 
parameters on the order of the unity,  and $a_S$ is the soft PQ breaking parameter, which allows the $S$ field to develop the vacuum expectation value on the order of $0.1-1$  TeV  in the present Universe. 
When the Higgs field develops the expectation values along the flat direction, $H_1\simeq H_2\sim \varphi \gg |m_S|$, 
$S$ becomes heavy with a mass of ${\cal O}(\varphi)$, and its vacuum value shifted by the $a_S$-term is quite suppressed as $\langle S \rangle \sim a_S^3/\varphi^2\ll m_0= \sqrt{(m_1^2+m_2^2)/2}$.
The resulting $b_H = \kappa S^2 \sim a_S^6/\varphi^4 $ is much smaller than $m_0^2$, and satisfies Eq.~(\ref{bcons}).   
As the $\varphi$ value decreases and becomes ${\cal O}(m_0)$, then $\langle S\rangle \sim m_0$, and $b_H \sim m_0^2$, 
so the PQWW axion becomes heavy with a mass of ${\cal O}(m_0)$, which is safe from various astrophysical/collider constraints.

We would like to emphasize that the scalar potential we suggest in this section is a proof of concept, in which a flat direction ($|H_1|=|H_2|$) exists and $b_\phi$-term 
is dynamical,
which is suitable for our magnetogenesis scenario. Clever ideas are welcome and desirable in order to provide a more natural set-up for our mechanism.  
See App.~\ref{HuHd} for a concrete example to realize the $H_1 H_2$ flat direction
without a bare $b_H$-term in a supersymmetric extension of the SM 
($H_1\to i\sigma_2 H_d^*$, and $H_2\to i\sigma_2 H_u^*$). 

\subsubsection{Effective action with light degrees of freedom}
Let us now see how the anomalous coupling $\sim (e^2/16 \pi^2)\theta F_{\mu\nu} {\tilde F}^{\mu\nu}$ is obtained in the low energy effective Lagrangian. 
Here we focus on the non-supersymmetric theory although we use the SUSY-inspired potential. 
When the Higgs fields obtain large field values along the flat direction $\varphi \gg m_0$, 
we can divide the fields, not only the Higgs field described in the above but also the matter and gauge fields, into heavy fields whose  masses are proportional to $\varphi$, and light fields which are massless or obtain masses at most with the soft breaking scales. 
The former includes the quarks, charged leptons, weak gauge bosons, and heavy Higgs fields, 
as well as the singlet scalar $S$, if any, 
and the latter includes the gluons, (electromagnetic) photons, neutrinos, and the light Higgs field (the \`a la AD field).
In the unitary gauge, 
the Lagrangian density for the light fields is 
\begin{align} \label{L_light}
-\frac{{\cal L}_{\rm light}}{\sqrt{-g}} =&~ \frac{1}{2} {\rm Tr} G_{\mu\nu} G^{\mu\nu} 
+ i \bar\nu_L  \sigma^\mu\partial_\mu\nu_L  
+ \frac{1}{4} F_{\mu\nu}F^{\mu\nu}  +\frac{1}{2}(\partial_\mu\varphi)^2 +
\frac{\varphi^2}{2}(\partial_\mu\theta)^2 \nonumber\\
&  + \frac{1}{2} (m_0^2- c_H H^2) \varphi^2 
+ \frac{|a_\phi|}{8M} \varphi^4 \cos(4\theta -\theta_A)  + \frac{\varphi^6}{8M^2},
\end{align} 
where $a_\phi=|a_\phi| e^{i\theta_A}$,  $m_0^2 = (m_1^2+ m_2^2)/2$, $c_H= -(c_1 + c_2)/2$.
$m_0^2$ can be naturally positive even if $m_1^2 m_2^2 <0 $ in order for electroweak symmetry breaking in the present Universe.  
The Lagrangian density for the heavy fields up to the quadratic order is given as
\begin{align}\label{L_heavy}
-\frac{{\cal L}_{\rm heavy}}{\sqrt{-g}}
 =&~ \frac{1}{2}(\partial_\mu H^0)^2  + \frac{1}{2}\Big(\frac{g^2\varphi^2}{4} \Big) (H^0)^2  + (\partial_\mu H^+)(\partial^\mu H^-)+ \Big(\frac{(g^2 + g'^2)\varphi^2}{4}  \Big) H^+ H^- \nonumber\\
& +\frac{1}{2} (\partial_\mu S_R)^2 + \frac{1}{2}\Big(m_S^2+ \frac{(\kappa_1+ \kappa_2 + \kappa)\varphi^2}{4}\Big) S_R^2 
+  \sqrt{2} a_S^3   S_R  \nonumber\\
&+  \frac{1}{2} (\partial_\mu S_I)^2  + \frac{1}{2} \Big(m_S^2+ 
 \frac{(\kappa_1+ \kappa_2 - \kappa)\varphi^2}{4}\Big) S_I^2 
 +\bar \psi_{ui} \Big(i\gamma^\mu D_\mu + \frac{y_{u i}}{2}\varphi e^{i \gamma_5\theta}\Big)\psi_{ui} \nonumber\\
 & 
+    \bar \psi_{di} \Big(i\gamma^\mu D_\mu + \frac{y_{d i}}{2}\varphi e^{i\gamma_5\theta}\Big)\psi_{di}
+   \bar \psi_{ei}\Big(i\gamma^\mu D_\mu + \frac{y_{ei}}{2}\varphi e^{i\gamma_5\theta}\Big)\psi_{ei} \nonumber\\
& + \frac{1}{2} W_{\mu\nu}^+ W^{\mu\nu -} 
+ \frac{g^2\varphi^2}{4} W_\mu^+ W^{\mu -}
+ \frac{1}{4} Z_{\mu\nu}^0 Z^{\mu\nu 0} 
+ \frac{(g^2 + g'^2)\varphi^2}{8} Z_\mu^0 Z^{\mu 0} , 
\end{align}
where Dirac fermions are constructed as
\begin{equation}
\psi_{ui }= \left(\begin{array}{c}  u_{L i} \\ u_{Ri }\end{array}\right), \
\psi_{di} = \left(\begin{array}{c}  d_{L i} \\ d_{Ri } \end{array}\right), \ 
\psi_{ei} = \left(\begin{array}{c}  e_{L i} \\  e_{Ri } \end{array}\right), \
\end{equation}
by using the chiral representation for the Dirac matrices, $S = (S_R + i S_I)/\sqrt{2}$, and for simplicity $\kappa$
and  $a_S$ are taken to be real. 
The unbroken gauge group is SU(3)$_{\rm C}\times$U(1)$_{\rm em}$, and the corresponding covariant derivative  is given by
\begin{equation}
D_\mu = \partial_\mu - i g_s T^a_\psi G^a_\mu  - i e q_{\psi} A_\mu,
\end{equation}
where $T^a_\psi$ is the generator  of SU(3)$_{\rm C}$,
 and $q_\psi$ is the EM charge for a given fermion $\psi$.
For quarks, $T^a_{u,d} = \lambda^a/2$, where $\lambda^a_{ij}$ are Gell-Mann matrices, and for charged leptons, $T^a_e=0$.  
The EM charges ($q_\psi$) are $q_{u}=2/3$, $q_d =-1/3$, and $ q_e = -1$.  
We ignore the interaction between $\theta$ and $S$ because it does not have any effect on our interest. 

For low energy scales much less than $\varphi$, the effective action can be obtained by integrating out heavy fields. 
Since the expectation values of heavy fields vanish, 
basically they do not leave any traces except anomalous couplings and threshold corrections.
While the latter can be absorbed by the redefinition of model parameters, 
the former should be added explicitly to the Lagrangian.
This is derived by calculating one-loop triangle diagrams mediated by heavy fermions ($\psi_{u_i}$, $\psi_{d_i}$, $\psi_{e_i}$) so that 
\begin{align}
- \frac{  {\cal L}_{\rm anom}}{\sqrt{-g}} =&~\frac{ 2 N_f g_s^2}{16\pi^2}  \theta {\rm Tr}G_{\mu\nu}\tilde G^{\mu\nu} 
+ \frac{N_f e^2}{16\pi^2} \left(3 q_u^2 + 3 q_d^2+ q_e^2 \right) \theta F_{\mu\nu} \tilde F^{\mu\nu} \nonumber\\
=&~ \frac{3 g_s^2 }{8\pi^2} \theta {\rm Tr}G_{\mu\nu}\tilde G^{\mu\nu} 
+ \frac{e^2}{2\pi^2} \theta F_{\mu\nu}\tilde F^{\mu\nu}.
\end{align}
Here $N_f$ is the number of heavy families, and we take $N_f=3$ for the SM. 
The appearance of such anomalous terms can be understood by noting that the flat direction is 
charged under PQ symmetry, which is anomalous under the SU(3)$_\mathrm{C}$ and  U(1)$_\mathrm{em}$, 
and all PQ charged fermions are heavy along the flat direction. Then, we arrive at the low energy effective Lagrangian density Eq.~\eqref{toylag}, and conclude that magnetogenesis is successful in the type-II 2HDM. 
We do not worry about the current induced by the Schwinger effect 
because all U(1)$_\mathrm{em}$ charged particles are massive.

\subsection{$L H_u$ flat direction in supersymmetric SM \label{sec:3B}}
In the previous section, we utilized some of the properties of the supersymmetric SM just to
justify a part of  the form of the scalar potential in the type-II 2HDM,  but did not take into account any SUSY partners. 
In this section,  we shall consider the supersymmetric extension of the SM more seriously, 
as is adopted in the AD mechanism. 
In the MSSM, or extended supersymmetric SMs, there are many scalar fields, namely, the SUSY partners of the SM fermions such as squarks and sleptons, 
which exhibit many flat directions~\cite{Gherghetta:1995dv}, 
along which the scalar potential vanishes except for the SUSY-breaking effects and contributions from non-renormalizable operators. 
Scalar fields can develop expectation values along a flat direction to cause the AD mechanism. 

As a proof of concept, let us focus on the $L H_u$ flat direction, which has been often used for the AD leptogenesis~\cite{Murayama:1993em}.
In order to make the scalar dynamics  simpler, we will consider   a flat direction only governed by a slepton with a single flavor $f$,  $\tilde L_{Lf}$, and $H_u$\footnote{
Note that multiple flavors of sleptons~\cite{Enqvist:2003pb} as well as the $H_d$ field~\cite{Kamada:2008sv} 
can co-exist with the $L_{Lf} H_u$ flat direction, which lead to possible multiple field dynamics in the AD mechanism. Indeed, in the MSSM, due to the $\mu$-term 
and $B\mu$-term, the expectation value of $H_d$ field is induced along the $L_{Lf} H_u$ flat direction. See App.~\ref{app:1} for the detail. },  
while $H_d$ and other scalar fields do not develop non-zero field values. 
Hereafter we use the tilde for supersymmetric partners.
Such a condition can be easily realized, {\it e.g.}, in the  next-to-minimal-supersymmetric Standard Model (NMSSM) with a superpotential, 
\begin{equation}
W_{\rm NMSSM} =  y_{u} Q_L u^c_R H_u + y_{d} Q_L d^c_R H_d + y_{e } L_L e^c_R H_d + \lambda S H_u H_d  
+ \frac{1}{2} m_S S^2 + \frac{1}{3}\kappa S^3. 
\end{equation} 
It can be easily seen that with the configuration 
\begin{equation}
(H_u)_{\textrm{$D$-flat}} = \left(\begin{array}{c}  0 \\ \dfrac{1}{2} \varphi e^{-i\theta} \end{array}\right), \ 
(\tilde L_{Lf})_{\textrm{$D$-flat}} =\left(\begin{array}{c}  \dfrac{1}{2} \varphi e^{-i\theta} \\  0\end{array}\right), 
\end{equation}  
the $D$-term potential as well as the $F$-term potential for the $\varphi$ and $\theta$ fields vanish and their potential is lifted only from the SUSY breaking effects as well as the Hubble induced terms
which give them ``small'' masses, on the order of the soft SUSY-breaking mass, $m_\mathrm{soft} = {\cal O}$(TeV), and the Hubble parameter, respectively. 
The scalar fields $H_d$ and $S$ acquire ``large'' masses of ${\cal O}(\lambda \varphi)$ along the $F$-flat direction so that we can integrate them out for  low energy effective theory.

Taking $H_d=S=0$ while keeping the $H_u$ and $\tilde L_{Lf}$ fields explicitly, the form of their scalar potential is the same as in Eq.~\eqref{V_AD} by replacing ($H_1,\, H_2$) to ($H_u,\, \tilde L_{Lf}$). 
Since $H_d$ and $L_L$ have the same SM gauge charges, this clearly shows that 
the expectation value of  $\varphi$  breaks  SU(2)$_L \times$ U(1)$_Y$ symmetry down to the U(1)$_\mathrm{em}$ so that the three scalar modes in the $H_u$ and $\tilde L_{Lf}$ fields  other than the $\varphi$ and $\theta$ fields are absorbed by vector bosons. 
Similarly, one CP even and one complex field also become also heavy  with the masses of ${\cal O}(g \varphi)$ from the $D$-term potentials.
As a result their field values can be safely set to be zero, and, again, they can be integrated out.
The low energy effective scalar potential along the $D$-flat direction, parameterized by the $\varphi$ and $\theta$ fields, is the same as that of Eq.~\eqref{L_light}.\footnote{The absence of other flavors of sleptons can be justified by supposing a positive Hubble induced mass for them.}

The  difference compared to the non-supersymmetric type-II 2HDM studied in the previous section 
is the additional fermionic degrees of freedom: Higgsinos ($\tilde H_u, \tilde H_d$) and 
gauginos ($\tilde W^a, \tilde B$),  and 
a pattern of the fermion mass splitting. 
While all the charged fermions get massive in the 2HDM case, massless charged fermions remain along the $L_{Lf} H_u$ flat direction. 
The Yukawa interactions are given by
\begin{equation}
\left(  y_{u i} H_u^0 u_{Li} u^c_{Ri}   + g (H_u^0)^*  \tilde W^- \tilde H_u^+   
  + y_{e f} \tilde L_{Lf}^0  \tilde H_d^- e_{Rf}^c  + g (\tilde L_{Lf}^0)^*  e_{Lf} \tilde  W^+  \right) + \mathrm{h.c.}, 
\end{equation}
 for the charged fermions who get masses from the expectation values of $H_u$ and $\tilde L_{Lf}$. 
In the unitary gauge, the corresponding Lagrangian density for the heavy fermions is written as 
\begin{align}
-\frac{{\cal L}_{\rm heavy}}{\sqrt{-g}} 
=&~ \bar \psi_{ui} \left( i \gamma^\mu D_\mu +  \frac{y_{u i}}{2} \varphi e^{ i \gamma_5\theta}\right) \psi_{u i} 
+    \bar\psi_{H_u} \left( i \gamma^\mu D_\mu + \frac{g}{2}\varphi e^{-i\gamma_5\theta}\right) \psi_{H_u} \nonumber\\
& +  \bar\psi_{H_d} \left( i \gamma^\mu D_\mu + \frac{y_{ef}}{2}\varphi e^{i\gamma_5\theta}\right) \psi_{H_d}
 + \bar\psi_W \left( i \gamma^\mu D_\mu + \frac{g}{2}\varphi e^{-i\gamma_5\theta}\right) \psi_W
\end{align}
where the Dirac fermions $\psi$ are defined as
\begin{equation}
\psi_{ui} = \left(\begin{array}{c}  u_{Li} \\  u_{Ri}^{c\dagger} \end{array}\right), \
\psi_{H_u} = \left(\begin{array}{c}  \tilde W^- \\ {\tilde H}_u^{+\dagger} \end{array}\right),\ 
\psi_{H_d} = \left(\begin{array}{c} \tilde H_d^- \\ e_{Rf}^{c\dagger}\end{array}\right), \
\psi_W = \left(\begin{array}{c}  e_{Lf} \\  {\tilde W}^{+\dagger}\end{array}\right).
\end{equation}
Note that $\psi_{H_d}$ and $\psi_W$ become heavy due to the non-zero $\langle \tilde L^f\rangle = \varphi/2$. 
They have the same electromagnetic charges ($q_e =-1 $), but couple to the axion oppositely, so
integrating them out does not yield low energy coupling between the axion and photons. 
This is consistent with the fact that lepton number is not anomalous under U(1)$_{\rm em}$.
There is no such cancellation between $\psi_{ui}$ and $\psi_{H_u}$ 
 ($q_u=2/3, q_{H_u} = q_e = -1$), providing the 
low energy couplings as  
\begin{align}
 -\frac{ {\cal L}_{\rm anom}}{\sqrt{-g}} = &~\frac{  N_f g_s^2}{16\pi^2}  \theta {\rm Tr}G_{\mu\nu}\tilde G^{\mu\nu} 
+ \frac{e^2}{16\pi^2} \left( N_f 3 q_u^2 -  q_e^2 \right) \theta F_{\mu\nu} \tilde F^{\mu\nu} \nonumber\\
=&~ \frac{3 g_s^2 }{16\pi^2} \theta {\rm Tr}G_{\mu\nu}\tilde G^{\mu\nu} 
+ \frac{3e^2}{16\pi^2} \theta F_{\mu\nu}\tilde F^{\mu\nu}.
\end{align}
Once more, we have used $N_f =3$. 
Since $\langle H_d\rangle=0$ during the evolution of  $\varphi$,  three $d$-quark pairs $\psi_{d i=1,2,3} =(d_{Li} \ d^{c \dagger}_{Ri})$, and two charged lepton pairs $\psi_{e i \not =f} =(e_{Li} \ e^{c \dagger}_{Ri})$ 
are massless. 
Because in our field basis those light charged fermions only couple to $H_d$, not $H_u$ and $\tilde L_f$,  there is no coupling between the axion and massless fermions.  
This can be seen by assigning U(1)$_{A'}$ charges to the fermion fields and the axion $\theta$ as 
\begin{equation}
Q_\psi\equiv -\frac{5}{4}B -\frac{7}{4} L  -  Y  -  q_{\rm em} +\frac{1}{4} q_{\rm PQ}, 
\end{equation}
so that the axion charge is one, $Q_\theta =1$, 
but the electromagnetic charged massless 
fermions are neutral. The PQ charge assignments $q_{\rm PQ}$  are given in 
Table~\ref{charge2}.
Since this U(1)$_{A'}$ contains the PQ charge, 
it is also anomalous 
under  SU(3)$_{\rm C}\times$U(1)$_{\rm em}$.

\begin{table}[h]
\begin{center}
\begin{tabular}{|c||c|c|c|c|c|c|c|c|c|c|}
\hline
 Fields& \, $Q_{Li}$\,   & \, $ u_{Ri}^c $ \, & \, $ d_{Ri}^c$\,  &\,  $ L_{Li} $\,  &\, $ e_{Ri}^c$\, &\,  $H_u $\,  &\, $H_d$\, &\, $S$ \,  \\
\hline
SU(2)$_L$ & ${\bf 2}$ & ${\bf 1}$ & ${\bf 1}$& ${\bf 2}$   &  ${\bf 1}$& ${\bf 2}$ & $ {\bf 2}$ & ${\bf 1}$ \\
\hline
U(1)$_Y$ & $1/6$ & $-2/3$ & $1/3$& $-1/2$   &  $1$& $1/2$ & $ -1/2$ & 0 \\
\hline
U(1)$_{PQ}$ & $1$ & $1$ & $1$& $1$   &  $1$& $-2$ & $ -2$ & 4 \\
 \hline
\end{tabular}
\end{center}
\caption{The SU(2)$_L$$\times$ U(1)$_Y$ and PQ charge assignment in the NMSSM}
\label{charge2}
\end{table}

By supposing higher dimensional operators 
\begin{equation}
W_{\rm NM} = \frac{(L_{Lf} H_u)^2}{M_f} + \cdots,
\end{equation}
and the negative Hubble induced quadratic terms for ${\tilde L}_f$ and $H_u$ in the same way as the 2HDM case, 
the final low energy Lagrangian density for the light fields is given by 
\begin{align} \label{L_light2}
-\frac{{\cal L}_{\rm eff}}{\sqrt{-g}} =&~ \frac{1}{2} {\rm Tr} G_{\mu\nu} G^{\mu\nu} 
+ i \bar\nu  \sigma^\mu\partial_\mu\nu
+ \frac{1}{4} F_{\mu\nu}F^{\mu\nu}  +\frac{1}{2}(\partial_\mu\varphi)^2 +
\frac{\varphi^2}{2}(\partial_\mu\theta)^2 \nonumber\\
&  + \frac{1}{2} (m_0^2- c_H H^2) \varphi^2 
+ \frac{|a_\phi|}{8M_f} \varphi^4 \cos(4\theta -\theta_A)  + \frac{\varphi^6}{8M_f^2} \nonumber\\
& 
+  \sum_{i=1}^3 \bar \psi_{di}  i\gamma^\mu D_\mu  \psi_{u i} +
  \sum_{i=1}^2  \bar \psi_{ei}  i\gamma^\mu D_\mu  \psi_{e i}  
  - \frac{  {\cal L}_{\rm anom}}{\sqrt{-g}}. 
\end{align} 
We have imposed the $a_\phi$-term while 
the $b_\phi$-term is absent since lepton number breaking is prohibited at the renormalizable level.
Thus we reach the effective Lagrangian in the form of Eq.~\eqref{toylag}, 
but massless U(1)$_\mathrm{em}$ charged particles also exist.

We can take a different field basis, by $\theta$ dependent chiral transformation of $d$-quarks and charged leptons. Then the axion photon couplings can be  removed 
through the chiral anomaly. Instead, axion-current interactions are generated.\footnote{
The coupling between the phase of the AD field and currents has been used for 
the realization of spontaneous baryogenesis in Refs.~\cite{Chiba:2003vp,Takahashi:2003db,Kamada:2012ht}. Our discussion suggests that magnetic fields are also produced 
in these setups.}  
Therefore this coupling is important for both the generation of gauge fields and the helicity of the fermions, 
which has also been discussed in the context of inflationary magnetogenesis in Ref.~\cite{Domcke:2018eki}.
Fermion production through the axion-current interaction has also been studied recently in Refs.~\cite{Adshead:2015kza,Adshead:2018oaa,Min:2018rxw,Adshead:2019aac}.

Because $W_{\rm NM}$ breaks lepton number and becomes the source of the neutrino masses as the Weinberg operator, 
there is the lower bound on $M$ 
from the upper bound on the neutrino masses $\sum m_\nu < {\cal O}(0.1 {\rm eV})$. 
On the other hand, 
since we do not know the lower bound on the lightest neutrino  mass, a very large value of $M_f$ is allowed. 
For example, in order to have the fiducial value for magnetogenesis, $\varphi_\mathrm{osc} \simeq 10^{12}$ GeV,
studied in Sec.~\ref{sec2}  for $m_0 \simeq 10^4$ GeV, $M_f \simeq 10^{20}$ GeV, 
we require a tiny neutrino mass $m_{\nu_f} \sim 10^{-7}$ eV. 

Let us comment about the effects of massless charged particles on magnetogenesis.
Through chiral anomaly, once helical magnetic fields are generated from the dynamics of the rotating scalar, 
fermions with chiral asymmetry will be also generated, by satisfying $\Delta h \simeq  (e^2/16 \pi^2) \Delta n_5$, 
with $n_5$ being the number density of the chiral  asymmetry. 
Moreover, through the Schwinger effect, non-chiral particles can be also generated, 
which can lead to thermalization of the charged particles~\cite{Tangarife:2017rgl}. 
As is discussed in Ref.~\cite{Domcke:2018eki}, these effects will suppress the efficiency of magnetogenesis. 
Thus we might not have as much magnetic helicity as much as evaluated in Sec.~\ref{sec2}. 

However, in the case of standard chiral plasma instability, the numerical MHD studies have shown that 
full transfer of the chiral asymmetry to the magnetic helicity is possible even in the fully thermalized system~\cite{Rogachevskii:2017uyc,Brandenburg:2017rcb,Schober:2017cdw,Schober:2018ojn}. 
From these observations, we expect that even in our case  the full transfer of the scalar asymmetry to the magnetic helicity can be accomplished in the existence of light particles as well as the thermal plasma.  For a concrete conclusion, nevertheless further investigation is needed, which is left for a future study. 

In this subsection we have focused on the $L H_u$ flat direction as a concrete example for proof of concept,  but we expect that similar effects can be seen in other flat directions in the supersymmetric SM, including the MSSM, because it is often the case that an unbroken U(1) gauge symmetry remains along the flat directions. 
For example, in the case of $u d d$ flat direction, a linear combination of the hyper gauge field, and the third and eighth gluons 
is unbroken and its anomalous coupling to the angular direction of the complex flat direction is expected.

In this section, we show that the new mechanism of magnetogenesis 
studied in Sec.~\ref{sec2} can be naturally realized in the PQWW axion dynamics as well as in the usual AD mechanism. 
As described in the introduction, our findings have two important messages. 
Namely, 1) by supposing a cosmic history like the AD mechanism, 
axions can generate magnetic fields efficiently. 2) In some cases, the AD mechanism also generates magnetic fields, which requires careful analysis of the scenario.
Since we have only studied some of simplified situations to show the proof of concept of the idea, 
further studies are needed to give precise and quantitative consequences of this effect.

\section{discussion \label{sec4}}

In this work, we studied the evolution of U(1) gauge fields that have an anomalous coupling
to the phase of a rotating complex scalar field, which is often realized in cosmology 
in the context of the AD mechanism. 
The existence of such an anomalous coupling is not surprising 
since the phase of the AD field can be identified as an axion. 
Compared to other existing scenarios of magnetogenesis from axion dynamics, 
in which the axion oscillates around the PQ breaking potential, 
our magnetogenesis is novel in the sense that the PQ breaking effects are important only at the onset of dynamics in the phase direction, and are absent during most of the evolution of the phase-field. 
As a result, only one helicity mode of the gauge fields is continuously subject to the tachyonic instability. 
This allows the full transfer of the asymmetry from the scalar field to magnetic helicity, making magnetogenesis more efficient. 
The mechanism studied in this work is analogous to the chiral plasma instability, 
in which the chiral magnetic effect induces an instability in the magnetic fields. 

We have shown that our mechanism can be realized in well-motivated extensions of the SM. As a proof of concept, we have demonstrated that the PQWW axion in the type-II 2HDM, as well as the phase of the complex $LH_u$ flat direction in the AD leptogenesis, can act as a phase-field of the rotating scalar in this new magnetogenesis scenario. We also note that magnetogenesis induced by anomalous couplings is a general phenomenon of the AD mechanism, which has not been recognized before. 

In order to evaluate the consequences of magnetogenesis, 
we employed a relatively simplified setup. 
Namely, we have assumed a negligible thermal plasma in the scalar field dynamics  and omitted the effects of possible light charged particles. 
The inclusion of the thermal effect would trigger an early onset of the scalar field rotation, making ${\dot \theta}$ vary during oscillations. 
The effect of light particles is the induction of an electric current, which corresponds to the Schwinger effect in a vacuum and an ohmic current in thermal plasma.  It will screen the electric field and suppress the efficiency of the magnetogenesis. Estimating the induced current in the presence of the chiral anomaly is a considerably involved task which we leave for future work. 
A natural consideration is whether the anomalous coupling of the AD field can play an important role at later times. In particular, one may expect that the coupling can introduce a new channel for $Q$-ball decay,  since this process breaks the global U(1) symmetry that guarantees the stability of $Q$ balls. 
However, while the size of a $Q$ ball is inversely related to the velocity of the phase-field,  the instability scale is larger than that by a factor of $1/\alpha$, where $\alpha$ is the fine structure constant. 
Therefore, at first glance, we expect that $Q$-ball decay triggered by anomalous coupling is not so  efficient, but this effect still could be interesting to explore in more depth.

\section*{Acknowledgments}

 We are grateful to Jeff Kost for careful reading of the manuscript. 
KK thanks IBS-CTPU for kind hospitality, where this work was initiated. 
The work of KK was supported by JSPS KAKENHI, Grant-in-Aid for Scientific Research JP19K03842 
and Grant-in-Aid  for Scientific Research on Innovative Areas 19H04610. 
KK and CS were supported by IBS under the project code, IBS-R018-D1.

\appendix

\section{$H_u H_d$ flat direction in supersymmetric SM }\label{HuHd}
In the MSSM, the $H_u H_d$ scalar field configuration is not a good flat direction for magnetogenesis, because the constant $B\mu$-term explicitly breaks the PQ symmetry at the renormalizable level. The corresponding quadratic scalar potential ($\Delta V \sim B\mu H_u H_d +h.c.$) strongly disturbs a long time rotation in the complex field space. 
In the NMSSM discussed in Sec.~\ref{sec:3B}, 
because of the quartic potential induced by the $F$-term ($\Delta V \sim \lambda^2 |H_uH_d|^2$), the $H_u H_d$ is not even a flat direction.
In this appendix, we construct a viable supersymmetric extension of the SM 
in which  the $H_u H_d$ field configuration exhibits a flat direction. 

Let us introduce two gauge singlet chiral superfields $S$ and $S^c$
with the PQ charge assignment in Table~\ref{charge3}.
Then the relevent superpotential can have the form
\begin{equation}
W = y_u Q_L u_R^c H_u + y_d Q_L d_R^c H_d + y_e L_L e_R^c H_d 
+ \frac{S^2 H_u H_d }{M_1}  +   \frac{S^3 S^c}{M_2} +  \frac{(H_u H_d)^2}{M_3},
\end{equation}
where 
$M_1$ $M_2$, and $M_3$ are very large constants compared to the weak scales.
\begin{table}[h]
\begin{center}
\begin{tabular}{|c||c|c|c|c|c|c|c|c|c|c|}
\hline
 Fields& \, $Q_{Li}$\,   & \, $ u_{Ri}^c $ \, & \, $ d_{Ri}^c$\,  &\,  $ L_{L i} $\,  &\, $ e_{R i}^c$\, &\,  $H_u $\,  &\, $H_d$\, &\, $S$ \, &\,  $S^c$\, \\
\hline
SU(2)$_L$ & ${\bf 2}$ & ${\bf 1}$ & ${\bf 1}$& ${\bf 2}$   &  ${\bf 1}$& ${\bf 2}$ & $ {\bf 2}$ & ${\bf 1}$ & ${\bf 1}$\\
\hline
U(1)$_Y$ & $1/6$ & $-2/3$ & $1/3$& $-1/2$   &  $1$& $1/2$ & $ -1/2$ & 0 & 0 \\
\hline
U(1)$_{PQ}$ & $1$ & $1$ & $1$& $1$   &  $1$& $-2$ & $ -2$ & 2 & $-6$\\
 \hline
\end{tabular}
\end{center}
\caption{The SU(2)$_L$$\times$ U(1)$_Y$ and PQ charge assignment}
\label{charge3}
\end{table}
The bare $\mu$ and $B\mu$ terms are forbidden at the renormalizable level. However, 
we allow higher dimensional operators which explicitly break the PQ symmetry such as  $(H_u H_d)^2/M_3$.  Including soft SUSY-breaking terms, the scalar potential of the neutral scalar fields is given by
\begin{align}
V=&\,  \left( m_S^2 + \frac{4 |h_u^0 h_d^0|^2}{M_1^2} \right) |S|^2  + \frac{9|S|^6}{M_2^2} 
+ \left(m_{S^c}^2   + \frac{|S|^4}{M_2^2}\right) |S^c|^2 + \left[ \left(\frac{A_2}{M_2} S^3 \right) S^c + h.c.\right] 
\nonumber\\
&
+ \left[ \frac{A_1}{M_1} S^2 +  \frac{6 |S|^2 (SS^c)^*}{M_1 M_2} 
+ \frac{ 2(|h_u^0|^2+|h_d^0|^2) (S^2)^* }{M_1 M_3} \right] h_u^0 h_d^0 + h.c.  \nonumber\\
&+ \left( m_{H_d}^2 + \frac{|S|^4}{M_1^2} \right) |h_d^0|^2  
+ \left( m_{H_u}^2 + \frac{|S|^4}{M_1^2} \right) |h_u^0|^2 +\frac{A_3}{M_3} (h_u^0 h_d^0)^2    \nonumber\\
&+  \frac{4(|h_u^0|^2 + |h_d^0|^2 ) |h_u^0 h_d^0|^2}{M_3^2}+ \frac{g^2+ g'^2}{8} \left(|h_u^0|^2 - |h_d^0|^2\right)^2.
\end{align}
In the present Universe, 
large  vacuum values of $S$ and $S^c$ are developed by the following scalar potential which is relevant for the dynamics of $S$ and $S^c$, assuming that $m_S^2 <0$, $m_{S^c}^2>0$.
\begin{align}\label{Svev}
&V(S) = - |m_S|^2 |S|^2 + \frac{9 |S|^6}{M_2^2} +  m_{S^c}^2 |S^c|^2 + \frac{ 2A_2 \langle |S|^3\rangle \cos(3\theta_S +\theta_{S^c})}{M_2} |S^c|  + \cdots  \nonumber\\
& \Rightarrow \langle |S|\rangle = {\cal O}(\sqrt{|m_S| M_2}),\ 
\langle |S^c|\rangle \sim {\cal O}\left(\frac{A_2 |m_S|}{m_{S^c}^2} \langle |S|\rangle\right).
\end{align}
Here, the nonzero expectation values of the $S$ and $S^c$ are induced by the negative mass squared term of $S$.
The large value of $M_2$ ensures the large expectation values of these fields, 
which justify the omission of the Higgs fields in evaluating them. Then 
the expectation value of $S$ dynamically induces 
$\mu$-term and $B\mu$-term as 
\begin{equation}
\mu = \frac{\langle S^2\rangle}{M_1} \sim |m_S| \frac{M_2}{M_1}, \quad 
B\mu = A_1\mu \sim A_1 m_S \frac{M_2}{M_1}.
\end{equation}
We naturally assume $M_1\sim M_2$, so that $\mu \sim |m_S|={\cal O}(m_{\rm soft})$, $B\mu \sim A_1 |m_S| ={\cal O}(m_{\rm soft}^2)$
are realized in the present Universe. 

On the other hand, for the large expectation values of the Higgs fields as  $|h_u^0|= |h_d^0| =\varphi/2 \sim \sqrt{m_0 M_3} \gg m_{\rm soft}$ with the assumption of $M_1^2 \sim M_2^2 \ll M_3^2$, 
the large positive mass squared term of $S$ is induced by the scalar potential  
$ \varphi^4 |S|^2/M_1^2 \sim (M_3/M_1)^2 m_0^2 |S|^2 \gg |m_S^2||S|^2$. 
Consequently, the $S$ field is trapped at the origin during magnetogenesis. Since the $B\mu$-term is dynamically absent in this period, we get a sizable amount of  magnetic fields through a long time evolution of the $H_uH_d$ phase field.

\section{$H_d$ field configuration along the $L H_u$ flat direction in the MSSM \label{app:1}}
In this appendix, we study the configuration of $H_d$ field along the $L H_u$ flat direction in the MSSM, in which $\mu$ and $B\mu$-terms are given by constants. 
We show that $H_d$ field gets expectation value induced by the $LH_u$ flat direction, 
which makes the anomalous coupling of the phase direction of the $LH_u$ flat direction 
to the photon vanish.

We consider the following superpotential of the MSSM and lepton number violating higher dimensional operators,
\begin{equation}
W = y_u Q_L u^c_R H_u + y_d Q_L d^c_R H_d + y_e Q_L e^c_R H_d + \mu H_u H_d +  \frac{(L_{Lf} H_u)^2}{M}. 
\end{equation}
Keeping in mind that $LH_u$ flat direction and $H_u H_d$ flat direction can 
coexist~\cite{Gherghetta:1995dv,Kamada:2008sv}, let us parameterize 
the relevant scalar degrees of freedom along the $LH_u$ flat direction as~\cite{Kamada:2008sv} 
\begin{equation}
\tilde L_{Lf} =  \left(\begin{array}{c} \tilde\nu \\ 0\end{array}\right),\quad 
H_u =\left(\begin{array}{c} 0 \\ \tilde h_u^0 \end{array}\right),\quad 
H_d = \left(\begin{array}{c} \tilde h_d^0  \\ 0\end{array}\right). 
\end{equation}
Then the scalar potential from $D$-terms, $F$-terms,  soft breaking terms, and the Hubble induced mass term is 
\begin{align}
V =&~ (m_L^2 + c_L H^2) |\tilde \nu|^2 + (m_{H_u}^2+\mu^2 + c_u H^2) |h_u^0|^2 + (m_{H_d}^2 +c_d H^2) |h_d^0|^2 \nonumber\\
&+ \left( \frac{A_\nu}{M} \tilde \nu^2 h_u^{02}  +  B\mu  h_u^0 h_d^0  + \mathrm{h.c.} \right)  
+ \frac{ 4|h_u^0|^4 |\tilde \nu|^2}{M^2} 
+ \left|\frac{2\tilde \nu^2 h_u^0}{M} + \mu h_d^0\right|^2 
\nonumber\\
&
+ \frac{g^2 + g'^2}{8} \left( |h_u^0|^2 - |h_d^0|^2 - |\tilde \nu|^2\right)^2, 
\end{align}
where $|c_u|, |c_d|, |c_L| = {\cal O}(1)$. We can easily see that the field configuration of the pure $L H_u$ flat direction ($|h_u^0|=|\tilde \nu|$) with  $|h_d^0|=0$ is impossible since 
there is the  tadpole potential for $h_d^0$, 
induced by SUSY breaking ($B\mu\langle h_u^0\rangle h_d^0$) and supersymmetric ($\mu \langle \tilde\nu^2 h_u^0\rangle^*h_d^0/M$) contributions. Therefore we have to estimate how $\langle h_d\rangle$ can be large along the $LH_u$ direction.

For the large values of $|h_u^0|$ and $|\tilde \nu|$ compared to soft SUSY breaking masses, the $D$-term potential makes one scalar degree heavy, so we can integrate out the corresponding field through the equations of motion.  
Parameterizing the scalar field amplitudes as 
\begin{equation}
|h_u^0| =  \varphi_u,\quad 
|h_d^0| =  \varphi_d, \quad |\tilde\nu|= \varphi_l.
\end{equation}
for $m^2, H^2 \ll  \varphi_d^2 \ll \varphi_u^2 \sim \varphi_l^2  \ll M^2$, which is realized for the negative Hubble induced mass
for $L$ and $H_u$,  $\varphi_l$ is determined by the $D$-flat condition, 
\begin{equation}
\varphi_l^2 \simeq \varphi_u^2 - \varphi_d^2 . 
\end{equation}
By imposing this $D$-flat condition, the potential for $\varphi_u$ and $\varphi_d$
as well as 
the gauge invariant phase fields, $\theta_{H}$ and $\theta_{L}$, defined as 
\begin{equation}
h_u^0 h_d^0 = \varphi_u\varphi_d e^{-i \theta_H} ,\quad 
\tilde\nu h_u^0 = \varphi_u \varphi_l e^{-i\theta_L}, 
\end{equation} 
is given by 
\begin{align}
V_{\rm eff}(\varphi_u, \varphi_d) =&~
\left[ m_{H_d}^2  +\mu^2 - m_L^2+  (c_d - c_L) H^2 \right] \varphi_d^2   \nonumber\\
& +\left[ 2 B\mu \varphi_d \cos \theta_H  -  \frac{4 \mu\varphi_d^3\cos (\theta_H- 2 \theta_L)}{M}  \right] \varphi_u  \nonumber\\
&+\left[ m_L^2 + m_{H_u}^2 + \mu^2 + (c_L+c_u) H^2 
 - \frac{2 A_\nu \varphi_d^2\cos ( 2\theta_L) }{M}+ \frac{4 \varphi_d^4}{M^2} \right] \varphi_u^2   \nonumber\\ 
&+\left[ \frac{4\mu \varphi_d \cos(\theta_H- 2\theta_L)}{M} \right] \varphi_u^3
 + \left[\frac{2 A_\nu \cos 2\theta_L}{M}  -  \frac{12 \varphi_d^2}{M^2}  \right] \varphi_u^4 
+ \frac{8 \varphi_u^6}{M^2}.
\end{align}
Here we have assumed that all constant parameters are real for simplicity.  

For $m\ll H$, with a reasonable assumption:
\begin{align}
m_L^2 +m_{H_u}^2 + \mu^2 + (c_L  + c_u ) H^2  <0, 
\end{align}
$\varphi_u$ gets a finite vacuum value as 
\begin{align}
V(\varphi_u) \sim ( c_L + c_u) H^2 \varphi_u^2 + \frac{8}{M^2} \varphi_u^6  
\ \Rightarrow    
\langle \varphi_u\rangle   = c \sqrt{HM},  
\end{align}
with  $c={\cal O}(1)$ whereas $\varphi_d \ll \varphi_u$. 
By inserting this to the potential, supposing $c_d-c_L-12 c^4>0$, 
the dominant contribution for the vacuum value of $\varphi_d$ is 
given by 
\begin{equation}
V(\varphi_d) \sim (c_d - c_L - 12 c^4) H^2\varphi_d^2 
+  \frac{4\mu \langle \varphi_u^3\rangle}{M} 
\cos(\theta_H - 2\theta_L)  \varphi_d
\ \Rightarrow \langle\varphi_d\rangle ={\cal O}\left(\frac{\mu}{H} \langle \varphi_u\rangle\right). 
\end{equation}
Note that the contribution from the $\mu$-term is stronger than that from the $B\mu$-term.
The angular field, $\theta_H$ also get 
a mass squared of ${\cal O}(\mu H \langle\varphi_u\rangle/\langle\varphi_d\rangle)\sim H^2$, 
so that $\theta_H$ is also heavy and follows the slow-rolling $\theta_L$ as $\langle\theta_H\rangle =2\theta_L + \pi$. 

As $H$ decreases and crosses the value of ${\cal O}(\mu)$,  
the field value of $\varphi_u$ becomes around $\sqrt{\mu M}$.  
Then the contribution of $B\mu$-term is no longer negligible for the potential of the 
$\varphi_d$  field so that
\begin{equation}
V(\phi_d)\sim 
(m_{H_d}^2 +\mu^2 - m_L^2)^2 \varphi_d^2 - (B\mu \varphi_u(t)  \cos \theta_H)\varphi_d  \ \Rightarrow 
\langle\varphi_d\rangle \sim \varphi_u(t). 
\end{equation}
Now the dynamics of $\theta_H$ is governed by the $B\mu$-term, 
which gives a constant heavy mass of ${\cal O}(\sqrt{B\mu})$, so  $\theta_H$ will exhibit the damped oscillation around $\pi$.
Therefore, while the phase of $LH_u$ rotates in the same way as the usual AD leptogenesis, 
the $H_uH_d$ rotation will  be quickly damped away. 
Since all the massless electromagnetic charged fermions in the pure 
$L H_u$ flat direction, such as $d$ quarks, 
acquire heavy masses from the $H_d$ field value, 
the anomalous coupling between the phase of $L H_u$ flat direction and photons 
is cancelled in the low energy effective theory.
Now we have found that the dynamical phase $\theta_L$ does not have the anomalous
coupling to photons and another phase $\theta_H$,  which has the anomalous coupling, 
no longer shows the constant velocity, 
we conclude that in the MSSM with a bare $B\mu$-term 
the magnetogenesis does not happen unless the $B\mu$-term is sufficiently suppressed 
as discussed in Sec.~\ref{sec:2d}. 
Note that in Ref.~\cite{Kamada:2008sv} the $B\mu$-term is not taken into account.  
This is the reason why ${\dot \theta_H}$ becomes constant and is not damped after the onset of 
scalar field oscillations around the origin there.

\end{document}